\documentclass[aps,reprint,superscriptaddress,longbibliography, prb]{revtex4-2}

\usepackage{graphicx}
\usepackage{amsmath}
\usepackage{bbm}
\usepackage{bm}
\usepackage{esint}
\usepackage{cancel}
\usepackage[dvipsnames]{xcolor}
\usepackage[unicode=true, colorlinks=true, citecolor={blue!80!black}, urlcolor={blue!50!black}, linkcolor = {blue!80!black}]{hyperref}
\usepackage{microtype}
\usepackage[normalem]{ulem}
\usepackage{mathrsfs}
\usepackage{multirow}
\usepackage{array}
\newcolumntype{C}[1]{>{\centering\arraybackslash}p{#1}}

\DeclareMathOperator{\sgn}{sgn}

\newcommand{\abs}[1]{\left\vert#1\right\vert}

\graphicspath{{../figures/}}

\begin{document}
\title{Quantum critical dynamics of a Josephson junction at the topological transition}

\author{Vladislav D.~Kurilovich}
\affiliation{Department of Physics, Yale University, New Haven, CT 06520, USA}

\author{Chaitanya Murthy}
\affiliation{Department of Physics, Stanford University, Stanford, CA 94305, USA}

\author{Pavel D.~Kurilovich}
\affiliation{Department of Physics, Yale University, New Haven, CT 06520, USA}

\author{Bernard van Heck}
\affiliation{Microsoft Quantum, Station Q, University of California, Santa Barbara, CA 93106, USA}
\affiliation{Microsoft Quantum Lab Delft, Delft University of Technology, 2600 GA Delft, The Netherlands}

\author{Leonid I.~Glazman}
\affiliation{Department of Physics, Yale University, New Haven, CT 06520, USA}

\author{Chetan Nayak}
\affiliation{Department of Physics, University of California, Santa Barbara, CA 93106, USA}
\affiliation{Microsoft Quantum, Station Q, University of California, Santa Barbara, CA 93106, USA}

\date{\today}
\begin{abstract}
We find the admittance $Y(\omega)$ of a Josephson junction at or near a topological transition. The dependence of the admittance on frequency and temperature at the critical point is universal and determined by the symmetries of the system. Despite the absence of a spectral gap at the transition, the dissipative response may remain weak at low energies: $\mathrm{Re}\,Y(\omega)\propto \max (\omega, T)^2$. This behavior is strikingly different from the electromagnetic response of a normal metal. Away from the critical point, the scaling functions for the dependence of the admittance on frequency and temperature are controlled by at most two parameters.
\end{abstract}

\maketitle

\section{Introduction\label{sec:intro}}

Superconducting quantum wires with broken time-reversal and spin-rotation symmetries recently emerged as a basis for a topological qubit \cite{kitaev2001, lutchyn2010, oreg2010}. By varying the applied magnetic field and chemical potential such wires may be tuned through a quantum critical point into a topological phase in which Majorana zero modes (MZMs) are localized at the ends of the wire. Two MZMs constitute a single non-local fermionic mode at zero energy that may be used to store quantum information in an intrinsically protected way. Non-Abelian exchange statistics of MZMs allows one to manipulate the information encoded in the degenerate ground state by performing braiding operations on the MZMs \cite{read2000, ivanov2001}. This lays a foundation for the field of topological quantum computation \cite{nayak2008}.

A necessary preliminary step towards topological quantum computation with superconducting quantum wires is a reliable detection of the topological phase. DC charge transport measurements in proximitized semiconducting nanowires reveal signatures consistent with the presence of MZMs, such as zero-bias conductance peaks \cite{mourik2012}. However, these measurements do not identify the topological phase unambiguously. Indeed, it was recently shown that local transport properties of MZMs can be mimicked by non-topological Andreev bound states \cite{liu2017, reeg2018, vuik2019}. Furthermore, other mechanisms (\textit{e.g.}, Kondo resonances \cite{lee2012} or disorder \cite{liu2012}) can provide alternative explanations for the observed zero-bias peaks. Thus the topological phase has so far remained elusive in DC experiments. This prompts exploration of complementary approaches to the identification of the topological phase.

A promising such approach is to use AC measurements. It has been predicted that the MZMs fused at the Josephson junction between two topological wires strongly modify the microwave response of the junction \cite{vayrynen2015, peng2016}, making it strikingly different from that of a conventional weak link in a trivial state \cite{kos2013}.  Earlier theoretical works studying microwave response mostly focused on systems deep in the topological phase \cite{ginossar2014, vayrynen2015, peng2016}. On the other hand, microwave experiments with superconducting wires have just started and so far have been limited to the investigation of the AC response of the trivial phase \cite{vanwoerkom2016, hays2018, tosi2019, hays2020, metzger2021}. This naturally leads to a question: what is the microwave response of a Josephson junction at and in the vicinity of the quantum critical point separating trivial and topological phases? With the notable exception of Ref.~\cite{tewari2012}, this question has received little theoretical attention. In this work, we provide its comprehensive study.

The critical dynamics of a Josephson junction at the topological phase transition is also of fundamental interest from the broader perspective of quantum-critical phenomena. The spectral gap closes at the critical point and the correlation-length and -time diverge. Such divergences are associated with power-law behavior of various response functions that may be characterized by a set of universal critical exponents \cite{sachdev2011}. We demonstrate how this paradigm is reflected in the critical behavior of the frequency-dependent admittance of the topological junction, $Y(\omega)$. This simple and experimentally accessible quantity characterizes the electromagnetic linear response of the junction.

\begin{figure}[t]
  \begin{center}
    \includegraphics[scale = 1]{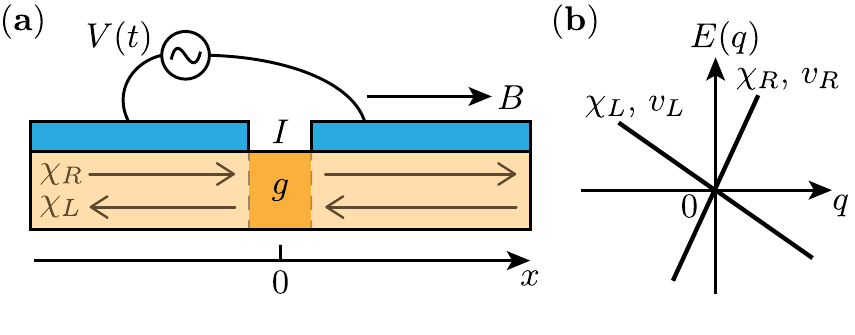}
    \caption{
    (a) Schematic layout of a considered setup. A Josephson junction (orange) is formed between two sections of a quantum wire (yellow) proximitized by superconducting shells (blue). An applied external magnetic field $B$ tunes the wire to the critical point. (b) The low-energy degrees of freedom in the wire at the critical point are a pair of counter-propagating Majorana modes $\chi_R$ and $\chi_L$ with linear dispersion. The Majorana modes are scattered at the junction [see Eq.~\eqref{eq:scat} for scattering parameter $g$]. We study the response of the current through the junction, $I$, to the applied alternating bias $V(t)$.}
    \label{fig:setup}
  \end{center}
\end{figure}

To find the admittance, we develop a universal theory that describes the low-energy degrees of freedom in the quantum wire at the critical point [see Sec.~\ref{sec:model}]. These degrees of freedom are a pair of counter-propagating Majorana modes with linear dispersion [see Fig.~\ref{fig:setup}]. Starting from this premise we show that, consistent with the notion of quantum criticality, the dissipative component of the junction's admittance depends on frequency $\omega$ and temperature $T$ as a power law, $\mathrm{Re}\,Y(\omega) \propto \max (\omega, T)^\gamma$ [see Sec.~\ref{sec:at}]. The dynamic critical exponent $\gamma$ is agnostic to the microscopic details of the system and depends only on its symmetries. We classify the critical theories of topological junctions by the presence or absence of two symmetries: (i) mirror reflection ${\cal M}_x$ with respect to a plane perpendicular to the wire that passes through the junction; and (ii) an antiunitary symmetry $\cal{R}$ that corresponds---on a microscopic level---to a combination of time-reversal and mirror reflection with respect to a plane containing the wire's axis [$x$-axis in Fig.~\ref{fig:setup}]. When at least one of the two symmetries is present, $\gamma = 2$ and the dissipative response depends strongly on frequency and temperature; dissipation becomes weak at small $\omega$ and $T$. This behavior is striking as the low-energy density of states is finite at the critical point. It occurs because the symmetries restrict the coupling of the critical modes to the electromagnetic field. The suppressed dissipative response highlights a fundamental difference between a critical quantum wire and a normal metal. Only when ${\cal M}_x$ and $\cal{R}$ are both absent do we find $\gamma = 0$, in which case $\mathrm{Re}\,Y(\omega)$ remains finite in the limit $\omega,\,T \rightarrow 0$, similarly to the dissipative conductance of a normal metal.

Upon detuning the system from the critical point a gap $E_\mathrm{gap}$ opens in the spectrum. The detuning can be acheived, \textit{e.g.}, by changing the external magnetic field $B$ applied to the wire from its critical value $B_\mathrm{c}$, in which case $E_\mathrm{gap} \propto |B - B_\mathrm{c}|$. The dissipative part of the admittance exhibits a scaling behavior with respect to $\omega$, $T$, and $E_\mathrm{gap}$.  We establish the most general form of the scaling function for each of the four possible combinations of symmetries ${\cal M}_x$ and $\mathcal{R}$ [see Sec.~\ref{sec:away}]. An interesting result of our theory is a pronounced asymmetry in the scaling behavior on the two sides of the topological transition. The asymmetry originates from the presence of an in-gap state that is localized at the junction and appears exclusively on one side of the transition \footnote{Depending on the microscopic details, the in-gap state may appear either in the trivial or in the topological phase.}.

We also investigate the non-dissipative component of the admittance, $\mathrm{Im}\,Y(\omega)$, across the topological phase transition [see Sec.~\ref{sec:non_diss}]. We show that it exhibits critical behavior as well, as manifested by the presence of a contribution $\propto (B - B_\mathrm{c}) \ln (B_\mathrm{c} / |B - B_\mathrm{c}|)$ that depends on $B - B_\mathrm{c}$ in a nonanalytic way.

Finally, we illustrate our universal theory with a particular model of a topological junction based on a proximitized semiconducting nanowire with strong Rashba spin-orbit coupling [see Sec.~\ref{sec:micro}]. The predictions of our theory can be directly tested in experiments in circuit quantum electrodynamics settings [see Sec.~\ref{sec:signatures} for a discussion].

\section{Model\label{sec:model}}
We start by introducing an effective field theory that describes the low-energy degrees of freedom in the quantum wire at the topological transition. Exactly at the critical point, the spectral gap closes and there exist two counter-propagating Majorana modes with linear energy dispersion. These helical modes are described at low-energies by the bulk Hamiltonian
\begin{equation}\label{eq:Hw}
    H_\mathrm{w} \approx -\frac{i}{2} \int dx \, \bigl(v_R \chi_R \partial_x \chi_R - v_L \chi_L \partial_x \chi_L\bigr),
\end{equation}
where $\chi_{R/L}(x)=\chi_{R/L}^\dagger(x)$ are Majorana field operators corresponding to right-moving ($R$) and left-moving ($L$) modes. The field operators satisfy anticommutation relations $\{\chi_i(x), \chi_j(x^\prime)\} = \delta_{ij}\delta (x-x^\prime)$. The propagation velocities $v_R$ and $v_L$ may be different in the general case.  Our model neglects the possible presence of disorder in the sample. A systematic study of the influence of disorder on the critical dynamics of a topological junction is left for future research.
 
The helical Majorana modes are scattered at the Josephson junction. The scattering is described by a local term in the Hamiltonian, $H_\mathrm{sc}$, that is bilinear in $\chi_{R/L}$. The expansion of $H_\mathrm{sc}$ in gradients of the fields generally starts with a contribution that has no derivatives:
\begin{equation}\label{eq:scat}
    H_\mathrm{sc} \approx -ig \, \chi_R(0) \chi_L(0),
\end{equation}
where $x = 0$ is the position of the junction. On a microscopic level, $H_\mathrm{sc}$ can originate from, \textit{e.g.}, imperfect transmission through the junction or a phase bias applied between the superconducting leads; see Sec.~\ref{sec:micro}. 

To study the electromagnetic response of the junction, it is necessary to establish the form of the current operator in the low-energy theory. In general, the current through the junction, $I\equiv I(x = 0)$, is a local operator which can be expressed as a bilinear form in the Majorana fields $\chi_{L/R}$ and their derivatives at $x=0$. To the first order in derivatives,
\begin{equation}\label{eq:current}
    I \approx  e\Big[i\alpha \chi_R(0)\chi_L(0) + i \kappa_{ij}\chi_i(0)\partial_x\chi_j(0)\Big],
\end{equation}
where $e > 0$ is the elementary charge, $\alpha$ and $\kappa_{ij}$ are real parameters, and a summation over $i,j = R,L$ is implicit in the second term. The necessity of keeping a subleading contribution in the gradient expansion of $I$ will become apparent momentarily.

The form of the critical theory defined by Eqs.~\eqref{eq:Hw}, \eqref{eq:scat}, and \eqref{eq:current} might be further constrained by the symmetries of the system. We focus on the implications of two fundamental discrete symmetries. The first one is a mirror reflection ${\cal M}_x$ with respect to the plane that is perpendicular to the wire and passes through the junction ($x = 0$). ${\cal M}_x$ acts on the Majorana fields in the following way:
\begin{equation}\label{eq:mirror}
      \chi_R(x) \overset{{\cal M}_x}{\rightarrow} \chi_L(-x),\quad \chi_L(x) \overset{{\cal M}_x}{\rightarrow} -\chi_R(-x).
\end{equation}
The second is an antiunitary symmetry $\cal{R}$ that exchanges right- and left-movers:
\begin{equation}\label{eq:antiunitary}
    \chi_R(x) \overset{\mathcal{R}}{\rightarrow} \chi_L(x),\quad  \chi_L(x) \overset{\mathcal{R}}{\rightarrow} \chi_R(x),\quad i \overset{\mathcal{R}}{\rightarrow} -i.
\end{equation}
Microscopically, $\cal{R}$ corresponds to a combination of time-reversal (which by itself is necessarily broken in a topological junction) and reflection with respect to a plane containing the $x$-axis along which the wire is oriented; see Sec.~\ref{sec:micro} and Appendix \ref{sec:app_R}.

\begin{table}
\begin{tabular}{ |C{1.90cm}|C{3.1cm}|C{3.1cm}|  }
 \hline
 &${\cal M}_x$ is present & ${\cal M}_x$ is absent\\
 \hline
 \multirow{2}{6em}{\centerline{$\cal{R}$ is present}} & $v_R = v_L,\,\alpha = 0,$   &$v_R = v_L,\,\alpha = 0,$\\
 &   $\kappa = \kappa_0 \mathbbm{1}$  & $\kappa = \kappa_0 \mathbbm{1} + \kappa_x \zeta_x$\\
 \hline
 \multirow{2}{6em}{\centerline{$\cal{R}$ is absent}} & $v_R = v_L,\,\alpha = 0,$   &  \multirow{2}{7.15em}{\centerline{no constraints}}\\
 &  $\kappa = \kappa_0 \mathbbm{1} + i\kappa_y \zeta_y$  & \\
 \hline
\end{tabular}
\caption{Classification of symmetry-imposed constraints on the low-energy theory [Eqs.~\eqref{eq:Hw}, \eqref{eq:scat}, and \eqref{eq:current}]. Here, $\mathbbm{1}$ is the $2 \times 2$ identity matrix and $\zeta_{x,y,z}$ are the Pauli matrices in the $R$/$L$ space, and $\kappa_{0,x,y}$ are real parameters.\label{tab:class}}
\end{table}

We begin the classification of symmetry-imposed constraints by noting that both ${\cal M}_x$ and $\cal{R}$ interchange right- and left-movers. Consequently, $v_R = v_L$ if either of these symmetries is present. Next, we note that, at the microscopic level, the current operator is odd under mirror reflection ${\cal M}_x$. If ${\cal M}_x$ is a symmetry of the system, this transformation property is retained upon projection to the low-energy subspace, \textit{i.e.}, the low-energy current $I$ transforms under ${\cal M}_x$ as $I\rightarrow -I$. The right hand side of Eq.~\eqref{eq:current} is consistent with this transformation law only if $\alpha = 0$ and $[\kappa, \zeta_y] = 0$, where $\zeta_{x,y,z}$ are the Pauli matrices in right-/left-mover space. Similarly, the microscopic current is odd under $\cal{R}$ (which incorporates time-reversal). Thus, if $\cal{R}$ is a symmetry of the system, the low-energy current must transform under $\cal{R}$ as $I \rightarrow -I$, which requires $\alpha = 0$ and $[\kappa, \zeta_x] = 0$. On the other hand, if {${\cal M}_x$ and ${\cal R}$ are not symmetries of the system,} then the transformation properties of the microscopic current need not be inherited by its low-energy counterpart, $I$. The reason is that the projection operator onto the low-energy subspace is not invariant under the symmetry transformation(s) in this case. Thus, in the absence of symmetries there are no constraints on the right hand side of Eq.~\eqref{eq:current}; in particular, $\alpha \neq 0$. The various symmetry-imposed constraints on the critical theory are summarized in Table~\ref{tab:class}.

It is instructive to determine how the four symmetry classes of Table~\ref{tab:class} fit into the general classification of topological matter \cite{altland1997, ryu2010, chiu2016}. Let us first note that ${\cal R}$ is an antiunitary symmetry satisfying ${\cal R}^2 = +1$. From the point of view of the Altland-Zirnbauer (AZ) classification \cite{altland1997}, this symmetry plays the role of an effective time-reversal operation. When ${\cal R}$ is present [upper row of Table \ref{tab:class}], the quantum wire belongs to class $\mathrm{BDI}$. When ${\cal R}$ is absent [bottom row of Table \ref{tab:class}], the wire has only the particle-hole symmetry of the AZ classification and therefore belongs to class $\mathrm{D}$. 
The possible presence of reflection symmetry ${\cal M}_x$ further enriches the topological classification \cite{chiu2016}. The case of ${\cal M}_x$ and ${\cal R}$ both present can be identified with the symmetry class $\mathrm{BDI}+R_{+-}$ of Ref.~\cite{chiu2016}.  The case of ${\cal M}_x$ present but ${\cal R}$ absent corresponds to the symmetry class $\mathrm{D}+R_-$. The topological invariants in the four symmetry classes of Table~\ref{tab:class} are listed in Ref.~\cite{chiu2016}. The invariants range from $\mathbbm{Z}_2$ in the simplest case of class $\mathrm{D}$ to $M\mathbbm{Z} \oplus \mathbbm{Z}$ in class $\mathrm{BDI}+R_{+-}$. The respective maximal  numbers of Majorana zero modes allowed by these invariants are vastly different from each other (interaction would limit the maximal number by $7$ \cite{fidkowski2011}). Yet, the symmetry classification of the quantum-critical behavior is captured by our Table~\ref{tab:class}, as long as only one additional Majorana zero mode emerges at the wire's end in a transition.

\section{Dissipative response at the critical point \label{sec:at}}

We now apply the low-energy theory of Sec.~\ref{sec:model} to study the dissipative component of the junction's admittance at the critical point. Let us assume that an alternating bias is applied to the junction [see Fig.~\ref{fig:setup}]. The influence of the bias on the system is described by a time-dependent perturbation of the Hamiltonian,
\begin{equation}\label{eq:drive}
    H_V \approx I\,\mathrm{Re} \bigl[i V e^{-i\omega t} / \omega \bigr],
\end{equation}
where $V$ is the bias amplitude \footnote{Notice that in Eq.~\eqref{eq:drive} we only present a part of the perturbation that acts on the low-energy degrees of freedom $\chi_{R/L}$. There might be other contributions to $H_V$ that involve high-energy degrees of freedom. These are irrelevant for the calculation of the dissipative response at small frequencies and temperatures, and are thus suppressed in Eq.~\eqref{eq:drive}.}. The admittance $Y(\omega)$ characterizes the response of the current $I$ to the perturbation $H_V$. It can be found at small drive strength ($e V \ll \omega$) via the linear response theory. The Kubo formula for the dissipative part of the admittance, $\mathrm{Re}\,Y(\omega)$, reads
\begin{equation}\label{eq:kubo}
    \mathrm{Re}\,Y(\omega) = -\frac{1}{\omega}\mathrm{Im}\, {\cal{C}}^R_{II}(\omega),
\end{equation}
with the response function given by
\begin{equation}\label{eq:resp}
    {\cal{C}}_{II}^R(\omega) = -i\int_0^{+\infty} dt \, e^{i\omega t} \langle [I(t), I(0)]\rangle.
\end{equation}
Here the average $\langle \cdots \rangle$ is performed over the Gibbs ensemble at temperature $T$. Physically, Eqs.~\eqref{eq:kubo}, \eqref{eq:resp} describe how drive photons are absorbed at the junction through processes in which either new Bogoliubov quasiparticles are produced or existing thermal ones are excited. 

The critical behavior of $\mathrm{Re}\,{Y}(\omega)$ can be deduced from scaling arguments. At the critical point, the dissipative admittance has a power law dependence on frequency and temperature.
The following estimate holds:
\begin{equation}\label{eq:critadm_gen}
    \mathrm{Re}\,Y(\omega) \sim \max(\omega, T)^\gamma,
\end{equation}
where $\gamma$ is a dynamic critical exponent. As follows from Eqs.~\eqref{eq:kubo}, \eqref{eq:resp}, $\gamma$ is related to the scaling dimension of the current operator in the critical theory, $[I]$:
\begin{equation}\label{eq:dynexp}
    \gamma = 2[I]-2.
\end{equation}
The scaling dimension $[I]$ depends on the symmetry of the system. If neither ${\cal M}_x$ nor $\cal{R}$ is present, the gradient expansion of $I$ starts with a term with no derivatives [see Eq.~\eqref{eq:current} and Table \ref{tab:class}]. Then, given that the Majorana fields have dimension $[\chi_{R/L}] = 1/2$, we find $[I] = 1$ and thus $\gamma = 0$. As a result, at low energies $\mathrm{Re}\,Y(\omega)$ does not depend on frequency and temperature,
\begin{equation}\label{eq:const}
    \mathrm{Re}\,Y(\omega) = c_\alpha = \mathrm{const}
\end{equation}
[here the constant $c_\alpha \propto \alpha^2$, with parameter $\alpha$ defined in Eq.~\eqref{eq:current}]. This behavior is similar to the dissipative conductance of a normal metal.

On the other hand, if at least one of ${\cal M}_x$ and $\cal{R}$ is a symmetry of the system, then $\alpha = 0$ and only gradient terms remain in Eq.~\eqref{eq:current}. As a result, the scaling dimension $[I] = 2$ and the critical exponent $\gamma = 2$. Consequently, in the presence of symmetries the dissipative part of the admittance depends strongly on frequency and temperature [see Eq.~\eqref{eq:critadm_gen}]. This is in spite of the constant density of states at the critical point. Specifically, we find that $\mathrm{Re}\,Y(\omega)$ can be represented as a combination of three terms (see Appendix \ref{sec:app_adm}):
\begin{subequations}\label{eq:critadm}
\begin{align}
    \mathrm{Re}\,Y_0(\omega) &=  c_0 \! \left[\omega^2 + (2\pi T)^2\right], \label{eq:critadm_0} \\
    \mathrm{Re}\,Y_y(\omega) &=  c_y \! \left[\omega^2 + (2\pi T)^2\right], \label{eq:critadm_y}\\
    \mathrm{Re}\,Y_x(\omega) &=  c_x \, \omega^2\label{eq:critadm_x} ,
\end{align}
\end{subequations}
with constants $c_j \propto \kappa_j^2$ [parameters $\kappa_j$ characterize different contributions to the current operator; they are introduced in Eq.~\eqref{eq:current} and Table~\ref{tab:class}]. The particular combination of terms is different for different symmetries, as summarized in Table~\ref{tab:answs}.

\begin{table}
\begin{tabular}{ |C{1.90cm}|C{3.1cm}|C{3.1cm}|  }
 \hline
 &${\cal M}_x$ is present &${\cal M}_x$ is absent\\
 \hline
 \multirow{2}{6em}{\centerline{$\cal{R}$ is present}} & $\gamma = 2$   &   $\gamma = 2$\\
 &   $\mathrm{Re}\,Y = \mathrm{Re}\,Y_0$  & $\mathrm{Re}\,Y = \mathrm{Re}\,Y_0 + \mathrm{Re}\,Y_x$\\
 \hline
 \multirow{2}{6em}{\centerline{$\cal{R}$ is absent}} & $\gamma = 2$   &  $\gamma = 0$\\
 &  $\mathrm{Re}\,Y = \mathrm{Re}\,Y_0 + \mathrm{Re}\,Y_y$ & $\mathrm{Re}\,Y = \mathrm{const}$\\
 \hline
\end{tabular}
\caption{Results for the dissipative component of the admittance at the critical point.
Here, $\gamma$ is the dynamic critical exponent [Eq.~\eqref{eq:dynexp}] and the functions $\mathrm{Re}\,Y_{0,y,x}$ are given by Eqs.~\eqref{eq:critadm_0}--\eqref{eq:critadm_x}.
\label{tab:answs}}
\end{table}

From Table \ref{tab:answs} and Eqs.~\eqref{eq:critadm_0}, \eqref{eq:critadm_y} it follows that
\begin{equation}\label{eq:resultIR}
\mathrm{Re}\,Y(\omega) = c_0\bigl[\omega^2 + (2\pi T)^2\bigr]
\end{equation}
 when ${\cal M}_x$ and $\cal{R}$ are simultaneously present, and
\begin{equation}\label{eq:resultInoR}
    \mathrm{Re}\,Y(\omega) = (c_0 + c_y) \bigl[\omega^2 + (2\pi T)^2 \bigr]
\end{equation}
when ${\cal M}_x$ is present but $\cal{R}$ is absent. Expressions~\eqref{eq:resultIR} and \eqref{eq:resultInoR} coincide up to a proportionality coefficient. This feature deserves explanation given that the current operator $I$ has different structure in the two cases [see Table~\ref{tab:class}]. 

Let us consider the case of ${\cal M}_x$ present and $\cal{R}$ absent. Invariance of the low-energy Hamiltonian under the $\mathbbm{Z}_2$ symmetry ${\cal M}_x$ actually leads to its invariance under the larger $U(1)$ group consisting of transformations
\begin{align}
\begin{aligned}\label{eq:transformation}
    \chi_R(x) &\rightarrow \chi_R(x) \cos (\vartheta / 2) + \chi_L(-x)  \sin (\vartheta / 2),\\
    \chi_L(x) &\rightarrow \chi_L(x) \cos (\vartheta / 2) - \chi_R(-x) \sin (\vartheta / 2) ,
\end{aligned}
\end{align}
with any real value of $\vartheta$ \footnote{We note that while the low-energy Hamiltonian with ${\cal M}_x$ present [Eqs.~\eqref{eq:Hw} and \eqref{eq:scat} with $v_R = v_L$] is invariant under transformation~\eqref{eq:transformation}, the latter does not correspond to any microscopic symmetry of the system. In other words, this is an emergent symmetry of the low-energy description.
The transformation~\eqref{eq:transformation} is formally the exponential of the mirror transformation~\eqref{eq:mirror} in the low-energy theory.}. At the same time, the current operator varies with $\vartheta$ upon applying the transformation~\eqref{eq:transformation}. By an appropriate choice of $\vartheta$ it can be brought to the form $I \propto i\chi_i(0)\partial_x \chi_i(0)$, identical to its form when both ${\cal M}_x$ and $\cal{R}$ are present. This implies that the admittances should coincide up to a numeric coefficient in the two cases depicted in the left column of Table \ref{tab:answs}.

Another notable feature of our results is a peculiar behavior of the contribution $\mathrm{Re}\,Y_x(\omega)$, which appears when $\cal{R}$ is present but ${\cal M}_x$ is absent. This contribution vanishes at $\omega = 0$ even if the temperature is finite [see Eq.~\eqref{eq:critadm_x}]. To explain this feature, we combine the Majorana fields into a Dirac fermion $\psi(x) = [\chi_R(x) + i \chi_L(-x)]/\sqrt{2}$. This fermion is chiral, $H_\mathrm{w} = -iv\int dx \, \psi^\dagger \partial_x \psi$. The scattering term $H_\mathrm{sc}$ in the Hamiltonian [see Eq.~\eqref{eq:scat}] can be represented through $\psi(x)$ as $H_\mathrm{sc} = g \psi^\dagger (0) \psi(0)$. Since $\psi(x)$ is chiral, $H_\mathrm{sc}$ cannot result in backscattering and merely leads to a phase shift. This phase shift may be eliminated with a gauge transformation; therefore, the eigenstates of the Hamiltonian are plane waves labeled by wave vector, $|k\rangle$. The term in the current operator that is responsible for contribution $\mathrm{Re}\,Y_x$ can be represented as $I_x = i\kappa_x (\chi_R \partial_x \chi_L + \chi_L \partial_x \chi_R) \propto \psi^\dagger \partial_x \psi - \psi \partial_x \psi^\dagger$ [its operator structure is not altered by the gauge transformation eliminating $H_\mathrm{sc}$]. The matrix element of $I_x$ between two plane waves is $\langle k |I_x |k^\prime \rangle \propto (k-k^\prime)$. In an absorption processes by a chiral fermion, energy conservation stipulates that $k - k^\prime \propto \omega$. Thus the matrix element vanishes at $\omega \rightarrow 0$, and $\mathrm{Re}\,Y_x(0) = 0$ regardless of temperature
\footnote{In the absence of scattering ($g = 0$), the fact that $\mathrm{Re}\,Y_x(0) = 0$ can be explained with an alternative---more physically transparent---argument. At $\omega \ll T$, $\mathrm{Re}\,Y_x(\omega)$ mainly stems from processes in which thermal quasiparticles absorb energy quantum $\omega$ from the drive $H_V$ and change their direction of motion. These processes become elastic in the limit $\omega \rightarrow 0$. However, the antiunitary symmetry $\cal{R}$ forbids elastic backscattering because it connects right-moving and left-moving states at a given energy:  $|L/R\rangle = \mathcal{R} |R/L\rangle$. Indeed, since ${\cal R} I {\cal R}^\dagger = -I$, the transition matrix element satisfies $\langle L| I |R\rangle = \langle R | {\cal R} I {\cal R}^\dagger | L \rangle^\star = -\langle L| I | R\rangle$ and thus vanishes. This is in direct analogy to how the elastic backscattering of edge modes is prohibited in time-reversal invariant topological insulators.}.

\begin{table}[]
\begin{tabular}{|C{2.025cm}|C{2.025cm}|C{2.025cm}|C{2.025cm}|}
\hline
$c_\alpha$ & $c_0$ & $c_x$ & $c_y$ \\ \hline
 \multirow{2}{6.8em}{\centerline{\hspace{-0.7em}$G_0 \dfrac{\alpha^2 \tau}{4v_R v_L}$}}    &
 \multirow{2}{6.8em}{\centerline{\hspace{-0.7em}$G_0 \dfrac{\kappa_0^2 \tau}{12v^4}$}}     &
 \multirow{2}{6.8em}{\centerline{\hspace{-0.7em}$G_0 \dfrac{\kappa_x^2 \tau}{4v^4}$}}      &
 \multirow{2}{6.8em}{\centerline{\hspace{-0.7em}$G_0 \dfrac{\kappa_y^2}{12 v^4}$}}\\
&&&\\
\hline
\end{tabular}
\caption{Explicit expressions for the proportionality constants in the equations for the dissipative component of the admittance [Eqs.~\eqref{eq:const}, \eqref{eq:critadm_0}--\eqref{eq:critadm_x}]. 
Here, $G_0 = e^2 / \pi$ is the conductance quantum and the parameter $\tau$ is given by Eq.~\eqref{eq:tau}.
In the expressions for $c_{0,x,y}$ the velocity $v = v_R = v_L$ [recall that the constants $c_{0,x,y}$ determine the admittance in the presence of ${\cal M}_x$ or ${\cal R}$, in which case $v_R = v_L$; see Table~\ref{tab:class}].
\label{tab:coef}}
\end{table}

The proportionality constants in Eqs.~\eqref{eq:const}, \eqref{eq:critadm_0}--\eqref{eq:critadm_x} can be explicitly related to the parameters of the low-energy theory (see Appendix~\ref{sec:app_adm} for details of the calculation). The relations are summarized in Table~\ref{tab:coef}, in which we introduced
\begin{equation}\label{eq:tau}
    \tau = \frac{1}{\cosh^2\!{\bigl[g/\sqrt{v_R v_L}\bigr]}}.
\end{equation}
This parameter has a physical meaning of transmission probability of the helical Majorana modes through the junction at the critical point.

\section{Dissipative response away from the critical point\label{sec:away}}

Away from the critical point [\textit{e.g.}, if the magnetic field $B$ is higher or lower than its critical value $B_\mathrm{c}$], right- and left-moving Majorana modes are hybridized. The hybridization is described at low energies by a mass term in the Hamiltonian:
\begin{equation}\label{eq:mass}
    \delta H_\mathrm{w} = -i M \int dx \, \chi_R(x) \chi_L(x),
\end{equation}
where the parameter $M$ characterizes the detuning from the critical point. We note that Eq.~\eqref{eq:mass} is applicable irrespective of which particular microscopic parameter is varied to tune the wire across the topological transition. Such a universality can be understood on the renormalization group grounds: $-i\int dx \chi_R(x)\chi_L(x)$ is the only RG-relevant homogeneous operator in the low-energy theory. Keeping the universality in mind, in what follows we assume for concreteness that the wire is tuned through the critical point by varying the magnetic field; in this case $M \propto B - B_\mathrm{c}$.

The mass term opens a gap $E_\mathrm{gap}$ in the energy spectrum of the system,
\begin{equation}\label{eq:gap}
    E_\mathrm{gap} = \frac{2\sqrt{v_R v_L}}{v_R + v_L} \abs{M} .
\end{equation}
In the presence of a gap, scattering at the junction [Eq.~\eqref{eq:scat}] might lead to the formation of a discrete \emph{non-degenerate} in-gap state. This happens on one side of the topological transition only: we find that the discrete state appears if $M\cdot g < 0$ (see Appendix~\ref{sec:app_ingap}). The energy $E_\tau$ of this state is given by
\begin{equation}\label{eq:bound}
   E_\tau = \sqrt{\tau}E_\mathrm{gap} ,
\end{equation}
where $\tau$ is the transmission probability of the helical modes through the junction at the critical point [see~Eq.~(\ref{eq:tau})].

\begin{figure}[t]
  \begin{center}
    \includegraphics[scale = 1]{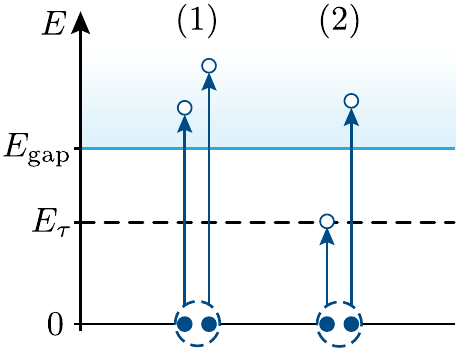}
    \caption{Two types of energy absorption processes that may contribute to $\mathrm{Re}\,Y(\omega)$ at $T = 0$. In processes of type $(1)$ a drive photon breaks a Cooper pair into two above-the-gap quasiparticles [$E > E_\mathrm{gap}$]. In processes of type $(2)$ a Cooper pair breaks into one quasiparticle at the discrete in-gap state, $E = E_\mathrm{\tau}$, and one quasiparticle above the continuum's edge, $E > E_\mathrm{gap}$.}
    \label{fig:processes_main}
  \end{center}
\end{figure}

The influence of the spectral gap on the frequency-dependence of the dissipative part of the admittance is most pronounced at low temperatures, $T \ll E_\tau \leq E_\mathrm{gap}$. Therefore, we focus here on the zero temperature limit [the results for the dissipative part of the admittance at finite temperature are presented in Appendix \ref{sec:app_adm}]. To compute $\mathrm{Re}\,Y(\omega)$ away from the critical point at $T = 0$, we represent the Kubo formula [Eq.~\eqref{eq:kubo}] as
\begin{align}
\label{eq:kuboT0}
\mathrm{Re}\,Y(\omega) = -\frac{2}{\pi\omega} &\int_0^\omega dE \ \mathrm{Tr}\Big\{ \hat{I} \bigl[G^R(-E) - G^A(-E)\bigr] \notag\\ &\times \hat{I} \bigl[G^R(\omega-E) - G^A(\omega-E)\bigr] \Big\} .
\end{align}
Here, $\hat{I}$ is a single-particle representation of the current operator \eqref{eq:current} and $G^{R/A}$ is the retarded/advanced Green's function of the Majorana modes (see Appendix \ref{sec:app_admittance} for details). The Schr\"{o}dinger equation for the Green's function is defined by the Hamiltonian $H = H_\mathrm{w} + \delta H_\mathrm{w} + H_\mathrm{sc}$ and can be solved exactly. This allows us to find a compact expression for $\mathrm{Re}\,Y(\omega)$ at arbitrary scattering strength $g$ in the presence of different combinations of symmetries.

Two types of processes may contribute to $\mathrm{Re}\,Y(\omega)$ at $T = 0$ [see Fig.~\ref{fig:processes_main}]. First, there are processes in which the energy of a drive photon is used to break a Cooper pair in the condensate and occupy two quasiparticle states above the continuum's edge. Such processes occur above the threshold frequency $\omega_\mathrm{th}^{(1)}=2E_\mathrm{gap}$. Formally, they correspond to a part of the integral in Eq.~\eqref{eq:kuboT0} in which both $E$ and $\omega - E$ are greater than $E_\mathrm{gap}$. We label the corresponding contribution to the dissipative part of the admittance as $\mathrm{Re}\,Y^{(1)}(\omega)$. Second, if $M\cdot g < 0$ the discrete state is present at the junction, and one of the two quasiparticles can occupy this state instead of going above the continuum's edge. These processes happen at frequencies $\omega>\omega_\mathrm{th}^\mathrm{(2)} = E_\mathrm{gap} + E_\tau$, and correspond to a part of the integral in Eq.~\eqref{eq:kuboT0} in which either $E$ or $\omega - E$ is less than $E_\mathrm{gap}$. We designate the respective contribution to the dissipative part of the admittance as $\mathrm{Re}\,Y^{(2)}(\omega)$. We note that the in-gap state \textit{does not} lead to a discrete line in the absorption spectrum. This is because two quasiparticles are produced in each absorption event whereas the in-gap state is non-degenerate and thus can accommodate only one quasiparticle. Below we separately study the contributions $\mathrm{Re}\,Y^{(1)}(\omega)$ and $\mathrm{Re}\,Y^{(2)}(\omega)$.

\subsection{Contribution $\mathrm{Re}\,Y^{(1)}(\omega)$}

\begin{table}
\begin{tabular}{ |C{1.90cm}|C{3.1cm}|C{3.1cm}|  }
 \hline
 &${\cal M}_x$ is present &${\cal M}_x$ is absent\\
 \hline
 \multirow{3}{6em}{\centerline{$\cal{R}$ is present}} & $\gamma = 2,\,\nu = 4$   &   $\gamma = 2,\,\nu = 2$\\
 &   $C = c_0$  & $C = c_0 + c_x$\\
 &  $|z(w,\varepsilon)|\rightarrow$ Eq.~\eqref{eq:matrixsymI} & $|z(w,\varepsilon)|\rightarrow$ Eq.~\eqref{eq:matrixsymnoI}\\
 \hline
 \multirow{3}{6em}{\centerline{$\cal{R}$ is absent}} & $\gamma = 2,\,\nu = 4$   &  $\gamma = 0,\,\nu = 2$\\
 &  $C = c_0 + c_y$ & $C = c_\alpha$\\
 &  $|z(w,\varepsilon)|\rightarrow$ Eq.~\eqref{eq:matrixsymI} & $|z(w,\varepsilon)|\rightarrow$ Eq.~\eqref{eq:matrixnosyms}\\
 \hline
\end{tabular}
\caption{Results for the dissipative part of the admittance away from the critical point. $C$ is a proportionality coefficient in the scaling relations \eqref{eq:scY1} and \eqref{eq:scY2}. It is determined by parameters $c_{\alpha,0,x,y}$ that were introduced in Eqs.~\eqref{eq:const}, \eqref{eq:critadm_0}--\eqref{eq:critadm_x}. $\gamma$ is the dynamic critical exponent. $\nu$ is another exponent that characterizes the behavior of the scaling function $f^{(1)}(w)$ near the absorption threshold [see Eq.~\eqref{eq:thr1} and related discussion]. $|z(w,\varepsilon)|$ is the transition matrix element that determines the scaling functions $f^{(1,2)}(w)$ [see Eqs.~\eqref{eq:scaling}, \eqref{eq:scaling2}].
\label{tab:answs_away}}
\end{table}

Away from the critical point, $\mathrm{Re}\,Y^{(1)}(\omega)$ assumes a scaling form:
\begin{equation}\label{eq:scY1}
    \mathrm{Re}\,Y^{(1)}(\omega) = \, C \omega^\gamma f^{(1)}\Bigl(\frac{\omega}{E_\mathrm{gap}}\Bigr).
\end{equation}
Here, $C$ is a constant [its relation to previously introduced coefficients $c_{\alpha,0,x,y}$ is presented in Table \ref{tab:answs_away}], $\gamma$ is the dynamic critical exponent [see Eqs.~\eqref{eq:critadm_gen}, \eqref{eq:dynexp}], and $f^{(1)}(w)$ is a dimensionless scaling function satisfying $f^{(1)}(w\rightarrow \infty) = 1$.
The scaling function can be expressed as (see Appendix \ref{sec:app_adm})
\begin{equation}\label{eq:scaling}
    f^{(1)}(w) = \frac{\Theta(w - 2)}{w} \! \int_{1}^{w-1} \!\! d\varepsilon\, \rho(\varepsilon)\rho(w-\varepsilon)|z(w, \varepsilon)|^2,
\end{equation}
where the step function highlights that production of a pair of quasiparticles in the continuum requires $\omega > \omega_\mathrm{th}^{(1)} = 2E_\mathrm{gap}$, $\varepsilon$ is the quasiparticle energy in units of $E_\mathrm{gap}$, $|z(w,\varepsilon)|$ is the transition matrix element, and $\rho(\varepsilon)$ is the local density of states (DOS) in the continuum at the junction [normalized by the density of states at the critical point, $(2\pi v_R)^{-1} + (2\pi v_L)^{-1}$]. In terms of dimensionless variables,
\begin{equation}\label{eq:ldos}
    \rho(\varepsilon) = \frac{\varepsilon\sqrt{\varepsilon^2 - 1}}{\varepsilon^2 - \tau}
\end{equation}
(see Appendix~\ref{sec:app_dos}). 
The expression for the matrix element $|z(w,\varepsilon)|$ depends on the symmetry of the system. If ${\cal M}_x$ and $\cal{R}$ are both absent, we find (see Appendix~\ref{sec:app_adm}):
\begin{equation}\label{eq:matrixnosyms}
    |z(w, \varepsilon)|^2 =  1-\frac{\tau}{(w-\varepsilon)\varepsilon}.
\end{equation}
If ${\cal M}_x$ is present, then the matrix element is given by
\begin{equation}
\label{eq:matrixsymI}
    |z(w, \varepsilon)|^2 = 3\Bigl(1-\frac{2\varepsilon}{w}\Bigr)^{2}\Bigl[1+\frac{\tau}{(w-\varepsilon)\varepsilon}\Bigr],
\end{equation}
regardless of whether $\cal{R}$ is present or not. As was shown in Sec.~\ref{sec:at}, in either case the current operator can be brought to the form $I \propto i[\chi_R(0)\partial_x \chi_R(0) + \chi_L(0)\partial_x \chi_L(0)]$ by a suitable unitary transformation [see Eq.~\eqref{eq:transformation} and the related discussion]. Notice that in both Eqs.~\eqref{eq:matrixnosyms} and \eqref{eq:matrixsymI}, the form of the matrix element is controlled by a single parameter, the transmission probability $\tau$. Consequently, the same holds for the respective scaling functions.

\begin{figure}[t]
  \begin{center}
    \includegraphics[scale = 1]{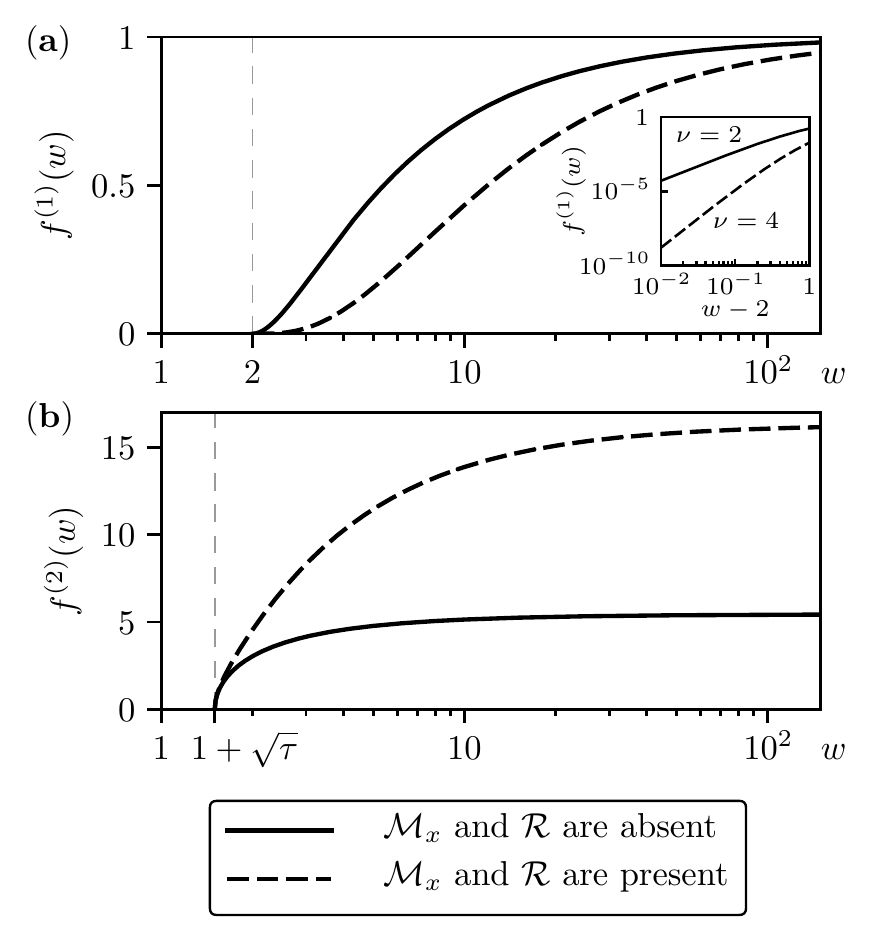}
    \caption{Examples of scaling functions for the dissipative part of the admittance obtained from Eqs.~\eqref{eq:scaling}, \eqref{eq:scaling2} for $\tau = 0.25$. Solid black curves demonstrate $f^{(1,2)}(w)$ in the case when ${\cal M}_x$ and $\cal{R}$ are both absent. Dashed black curves correspond to the case in which ${\cal M}_x$ and $\cal{R}$ are both present. Scaling function $f^{(1)}(w)$ [panel (a)] vanishes below the absorption threshold $w = 2$. Above the threshold, $f^{(1)}(w) \propto (w - 2)^\nu$ [see Eq.~\eqref{eq:thr1}], where $\nu = 2$ in the absence of ${\cal M}_x$ [solid curve] and $\nu = 4$ in the presence of ${\cal M}_x$ [dashed curve]. Scaling function $f^{(2)}(w)$ [panel (b)] is characterized by a lower absorption threshold, $w = 1 + \sqrt{\tau}$. Above the threshold $f^{(2)}(w)\propto (w - 1 - \sqrt{\tau})^{1/2}$ regardless of the symmetry of the system [see Eq.~\eqref{eq:thr2}]. Both $f^{(1)}(w)$ and $f^{(2)}(w)$ approach constant values as $w \to \infty$ [see the comment after Eq.~\eqref{eq:scY2}].}
    \label{fig:scaling}
  \end{center}
\end{figure}

Finally, if ${\cal M}_x$ is absent but $\cal{R}$ is present the matrix element is given by
\begin{align}\label{eq:matrixsymnoI}
    |z(w, \varepsilon)|^2 &= \,\lambda \cdot 3 \Bigl(1-\frac{2\varepsilon}{w}\Bigr)^{2}\Bigl[1+\frac{\tau}{(w-\varepsilon)\varepsilon}\Bigr]\notag\\
    &\quad\, + (1 - \lambda) \Bigl[1-\frac{\tau}{(w-\varepsilon)\varepsilon}\Bigr],
\end{align}
where $\lambda = c_{0}/(c_0 + c_x)$; the parameters $c_{0,x}\propto \kappa_{0,x}^2$ were introduced in Eqs.~\eqref{eq:critadm_0}, \eqref{eq:critadm_x} [also see Table \ref{tab:coef}]. Contrary to the other symmetry combinations considered, in this case the scaling function depends on an extra parameter $\lambda$, in addition to $\tau$.

Near the absorption threshold, $w - 2 \ll 1 - \tau$, the scaling function $f^{(1)}(w)$ depends on $w - 2$ as a power law,
\begin{equation}\label{eq:thr1}
    f^{(1)}(w) \propto \Theta (w - 2) \, (w - 2)^{\nu}.
\end{equation}
By analyzing Eq.~\eqref{eq:scaling}, we find that the exponent is $\nu = 4$ if ${\cal M}_x$ is present (regardless of the presence or absence of $\cal{R}$), and $\nu = 2$ if ${\cal M}_x$ is absent. The behavior of $f^{(1)}(w)$ is demonstrated in Fig.~\ref{fig:scaling}(a).

\subsection{Contribution $\mathrm{Re}\,Y^{(2)}(\omega)$}

The contribution to the dissipative admittance due to the discrete state, $\mathrm{Re}\,Y^{(2)}(\omega)$, can also be represented in a scaling form:
\begin{equation}\label{eq:scY2}
    \mathrm{Re}\,Y^{(2)}(\omega) = \, C E_\mathrm{gap}\,\omega^{\gamma - 1} f^{(2)}\Bigl(\frac{\omega}{E_\mathrm{gap}}\Bigr),
\end{equation}
where $C$ and $\gamma$ are the same parameters as in Eq.~\eqref{eq:scY1} and the scaling function satisfies $f^{(2)} (w \rightarrow \infty) = \mathrm{const}$ [note that since the factor $C$ is chosen to be the same in Eqs.~\eqref{eq:scY1} and \eqref{eq:scY2}, $f^{(2)} (w \rightarrow \infty) \neq 1$]. The scaling function is given by (see Appendix~\ref{sec:app_adm})
\begin{align}\label{eq:scaling2}
    f^{(2)}(w) &= 2\pi\,\Theta(w - 1-\sqrt{\tau}) \notag \\ 
    &\quad\, \times \sqrt{1-\tau} \, \rho(w-\sqrt{\tau})|z(w,\sqrt{\tau})|^2.
\end{align}
Here, the step function indicates that $\mathrm{Re}\,Y^{(2)}(\omega)$ is non-zero only above the absorption threshold $\omega_\mathrm{th}^{(2)} = E_\mathrm{gap} + E_\tau$. The local DOS in the continuum, $\rho(w - \sqrt{\tau})$, is given by Eq.~\eqref{eq:ldos}, and the factor $\sqrt{1 - \tau}$ originates from a subgap contribution to the local DOS due to the discrete state (see Appendix \ref{sec:app_dos}). This factor approaches zero for $\tau \rightarrow 1$ because the wavefunction of the discrete state becomes spatially extended in this limit. The transition matrix element $|z(w, \sqrt{\tau})|$ depends on the symmetry of the system. It is given by Eq.~\eqref{eq:matrixnosyms} if both ${\cal M}_x$ and $\cal{R}$ are absent, by Eq.~\eqref{eq:matrixsymI} if ${\cal M}_x$ is present, and by Eq.~\eqref{eq:matrixsymnoI} if $\cal{R}$ is present but ${\cal M}_x$ is absent. 

We note that, despite the fact that the matrix element is different for different combinations of symmetries, the power-law behavior of $f^{(2)}(w)$ near the absorption threshold is symmetry-independent. For  $w - 1 - \sqrt{\tau} \ll 1 - \tau$ we find
\begin{equation}\label{eq:thr2}
   f^{(2)}(w) \propto \Theta(w - 1 - \sqrt{\tau}) \, (w - 1 - \sqrt{\tau})^{1/2},
\end{equation}
as demonstrated in Fig.~\ref{fig:scaling}(b). This sharp square-root dependence mimics the local DOS above the edge of the continuum [see Eq.~\eqref{eq:ldos}].

Equations~\eqref{eq:scY1} and \eqref{eq:scY2} allow for the comparison of the two contributions to the dissipative part of the admittance. While $\mathrm{Re}\,Y^{(2)}(\omega)$ dominates over $\mathrm{Re}\,Y^{(1)}(\omega)$ close to $\omega^{(1)}_\mathrm{th}$, at high frequencies $\omega \gg E_\mathrm{gap}$ we estimate  $\mathrm{Re}\,Y^{(2)}(\omega)/\mathrm{Re}\,Y^{(1)}(\omega) \propto E_\mathrm{gap}/\omega \ll 1$.

\begin{figure}[t]
  \begin{center}
    \includegraphics[scale = 1]{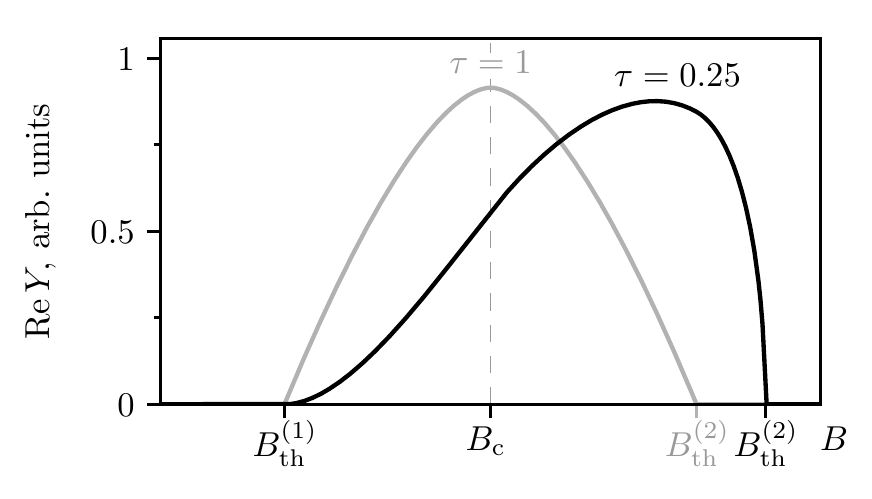}
    \caption{Dependence of the dissipative part of the admittance, $\mathrm{Re}\, Y$, on magnetic field $B$ in the vicinity of the critical point ($B = B_\mathrm{c}$) at a fixed frequency $\omega$. Black and gray curves correspond to $\tau = 0.25$ and $\tau = 1$, respectively [scales are different for the two curves]; they are plotted with the help of Eqs.~\eqref{eq:scY1} and \eqref{eq:scY2} in which we have varied $E_\mathrm{gap} \propto |B - B_\mathrm{c}|$ at fixed $\omega$. For concreteness, we assumed that ${\cal M}_x$ and $\cal{R}$ are both absent. The absorption thresholds $B_\mathrm{th}^{(1)}$ and $B_\mathrm{th}^{(2)}$ correspond to the values of the magnetic field at which $E_\mathrm{gap} = \omega / 2$ and $E_\mathrm{gap} = \omega / (1 + \sqrt{\tau})$, respectively. At $\tau = 0.25$ the dependence of $\mathrm{Re}\,Y$ on $B$ is asymmetric across the critical point: the curve is skewed towards the side of the topological transition at which there is a bound state at the junction. At $\tau = 1$ there is no bound state at either side of the transition and the curve is symmetric with respect to $B = B_\mathrm{c}$. We note that, up to a rescaling, the figure would look the same if another control parameter (\textit{e.g.}, the chemical potential) was used instead of $B$ to tune the wire across the topological transition.
    }
    \label{fig:rey}
  \end{center}
\end{figure}

Finally, Eqs.~\eqref{eq:scY1} and \eqref{eq:scY2} can be used to study the dependence of the dissipative part of the admittance on magnetic field near the critical point. We recall that in the low-energy theory the magnetic field enters the expression for the admittance via the parameter $M \propto B-B_\mathrm{c}$ which determines the spectral gap $E_\mathrm{gap}$ [see Eqs.~\eqref{eq:mass}, \eqref{eq:gap}]. An example of the magnetic field dependence of $\mathrm{Re}\,Y(\omega)$ at a fixed frequency $\omega$ is shown in Fig.~\ref{fig:rey}. The figure highlights that that $\mathrm{Re}\,Y(\omega)$ is an asymmetric function of $B - B_\mathrm{c}$. It has a peak that is displaced towards the side of the topological transition at which there is a bound state at the junction [\textit{i.e.}, the side at which the contribution $\mathrm{Re}\,Y^{(2)}(\omega)$ is present].

\section{Non-dissipative part of the admittance \label{sec:non_diss}}

Having discussed the dissipative part of the admittance in the vicinity of the critical point, we proceed to the analysis of the non-dissipative component $\mathrm{Im}\,Y(\omega)$. In this section we focus on the limit of $T = 0$ for simplicity.

We first address the behavior of $\mathrm{Im}\,Y(\omega)$ at the critical point. To this end, we note that $Y(\omega)$ is an analytic function in the upper-half complex plane of $\omega$. Then, because its real part, $\mathrm{Re}\,Y(\omega)$, is by itself an analytic function---either a constant or $\omega^2$ depending on symmetry [see Eqs.~\eqref{eq:const}, \eqref{eq:critadm}]---its imaginary part, $\mathrm{Im}\,Y(\omega)$, should also depend on $\omega$ in an analytic way. The precise form of this analytic dependence cannot be established on the basis of the low-energy theory alone since the high-energy modes in the wire can significantly contribute to $\mathrm{Im}\,Y(\omega)$. We note, however, that a general requirement of the linear response theory is that $\mathrm{Im}\,Y(\omega)$ is an odd function of $\omega$. This function scales as $1/\omega$ at sufficiently low frequencies because it includes the inductive response of the superconducting condensate.

While our low-energy theory cannot fully access the frequency dependence of $\mathrm{Im}\,Y(\omega)$ at the critical point, it can be applied to establish how $\mathrm{Im}\,Y(\omega)$ changes as a function of magnetic field $B$ in the vicinity of the topological transition, $|B - B_\mathrm{c}|\ll B_\mathrm{c}$. To do that, we consider the difference
\begin{equation}
    \delta \mathrm{Im}\,Y(\omega) = \left.\mathrm{Im}\,Y(\omega)\right|_B - \left.\mathrm{Im}\,Y(\omega)\right|_{B_\mathrm{c}},
\end{equation}
where $\left.\mathrm{Im}\,Y(\omega)\right|_B$ denotes the non-dissipative part of the admittance at external magnetic field $B$. As we show in Appendix \ref{sec:app_linresp}, $\delta \mathrm{Im}\, Y(\omega)$ is given by
\begin{equation}\label{eq:im_adm}
    \delta\mathrm{Im}\,Y(\omega) = \delta \mathrm{Im}\,Y_i(\omega) + \delta \mathrm{Im}\,Y_v(\omega).
\end{equation}
The first term describes the inductive response of the superconducting condensate,
\begin{equation}\label{eq:im_adm_ind}
    \delta\mathrm{Im}\,Y_i(\omega) = \frac{(2e)^2}{\omega}\partial^2_\varphi \bigl(\delta E_\mathrm{gs}\bigr),
\end{equation}
where $E_\mathrm{gs}\equiv E_\mathrm{gs}(\varphi, B)$ denotes the ground state energy of the system, $\varphi$ is the phase difference across the junction, and $\delta E_\mathrm{gs} = E_\mathrm{gs}(\varphi, B) - E_\mathrm{gs}(\varphi, B_\mathrm{c})$. The second term in Eq.~\eqref{eq:im_adm} can be obtained from $\mathrm{Re}\,Y(\omega)$ using the Kramers-Kronig relations:
\begin{equation}\label{eq:im_adm_v}
    \delta\mathrm{Im}\,Y_v(\omega) = -\frac{2\omega}{\pi} \fint_{0}^{+\infty}\frac{\delta\mathrm{Re}\,Y(\omega^{\prime})}{\omega^{\prime 2}-\omega^{2}}d\omega^{\prime}.
\end{equation}
Here, $\fint$ indicates the Cauchy principal value of the integral and $\delta \mathrm{Re}\,Y(\omega) = \left.\mathrm{Re}\,Y(\omega)\right|_B - \left.\mathrm{Re}\,Y(\omega)\right|_{B_\mathrm{c}}$. 
This contribution can be interpreted as originating from the virtual transitions that occur due to the applied bias.

We first analyze the inductive term $\delta \mathrm{Im}\,Y_i \propto \partial^2_\varphi (\delta E_\mathrm{gs})$. To do so, we note that at $|B - B_\mathrm{c}| \ll B_\mathrm{c}$ the leading contribution to $\delta E_\mathrm{gs}$ comes from the states of the quasiparticle continuum, whereas the bound state might produce a subleading correction only (as will be justified momentarily). The continuum contribution can be found as \cite{souma2002, murthy2020}
\begin{equation}\label{eq:deltaEgs}
    \delta E_\mathrm{gs} \approx  \delta\Bigl\{-\frac{1}{2} \int_{E_\mathrm{gap}}^{+\infty} \frac{EdE}{2\pi i} \frac{\partial}{\partial E}\ln \det S(E)\Bigr\},
\end{equation}
where $S(E)$ is the quasiparticle scattering matrix and $\delta$ on the right hand side denotes the difference between the expressions at $B$ and $B_\mathrm{c}$. The scattering matrix at relevant energies $E \sim E_\mathrm{gap}$ can be found using the low-energy theory; see Appendix~\ref{sec:app_scat_mat}. Its determinant is given by
\begin{equation}\label{eq:detS}
    \det S = \frac{\sqrt{E^2 - E_\mathrm{gap}^2} - i \sgn(M\cdot g)  E_\mathrm{gap} \sqrt{1 - \tau}}{\sqrt{E^2 - E_\mathrm{gap}^2}  + i \sgn(M\cdot g)  E_\mathrm{gap}\sqrt{1 - \tau}}.
\end{equation}
Using Eqs.~\eqref{eq:im_adm_ind}, \eqref{eq:deltaEgs}, and \eqref{eq:detS} we estimate $\delta\mathrm{Im}\,Y_i$ as
\begin{equation}\label{eq:cont_non_analyt}
    \delta \mathrm{Im}\,Y_i \approx -\frac{(2e)^2}{2\pi}\sgn(M\cdot g) \frac{E_\mathrm{gap}}{\omega} \ln \Bigl[\frac{\Delta}{E_\mathrm{gap}}\Bigr] \partial^2_\varphi \sqrt{1 - \tau},
\end{equation}
where we regularized the logarithmic divergence of the energy integral by an ultraviolet cut-off of the order of the proximity induced pairing scale $\Delta$, and neglected all subleading corrections. Equation \eqref{eq:cont_non_analyt} highlights the critical behavior of the non-dissipative admittance: $\delta \mathrm{Im}\,Y_i$ depends on $B - B_\mathrm{c}$ in a nonanalytic way, $\delta \mathrm{Im}\,Y_i \propto (B - B_\mathrm{c}) \ln (B_\mathrm{c}/ |B - B_\mathrm{c}|)$, due to the behavior of the gap across the transition [there is no absolute value in the factor outside the logarithm, due to the multiplier $\sgn(M\cdot g)$ in Eq.~\eqref{eq:cont_non_analyt}]. The factor $\partial^2_\varphi \sqrt{1 - \tau}$ in Eq.~\eqref{eq:cont_non_analyt} is sensitive to the microscopic details of the junction and cannot be determined on the basis of the low-energy theory alone. However, it is not critical and can be replaced by its value at $B = B_\mathrm{c}$.  Finally, we note that the bound state---which is present at the junction when $M\cdot g < 0$---also contributes to $\delta E_\mathrm{gs}$ and $\delta \mathrm{Im}\,Y_i$. Its contribution scales as $\propto \Theta(-M\cdot g)(B - B_\mathrm{c})$ and can thus be neglected in comparison with the logarithmically-larger contribution due to continuum states [see Eq.~\eqref{eq:cont_non_analyt}].

We now turn to $\delta\mathrm{Im}\,Y_v$ [see Eq.~\eqref{eq:im_adm}]. This contribution to the non-dissipative part of the admittance can be found analytically at $E_\mathrm{gap} \ll \omega\ll \Delta$ using the low-energy expressions for $\mathrm{Re}\,Y(\omega)$ (see Appendix~\ref{sec:app_imYv} for a detailed discussion). When ${\cal M}_x$ and ${\cal R}$ are both absent, we obtain
\begin{equation}\label{eq:imYv}
    \delta\mathrm{Im}\,Y_v \approx -2C\sqrt{1-\tau}\, \sgn(M\cdot g)\frac{E_\mathrm{gap}}{\omega}\ln \Bigl[\frac{\omega}{E_\mathrm{gap}}\Bigr].
\end{equation}
In this case, the dependence of $\delta\mathrm{Im}\,Y_v$ on $B - B_\mathrm{c}$ has a logarithmic feature similar to that in the inductive contribution $\delta \mathrm{Im}\,Y_i$ [cf.~Eqs.~\eqref{eq:cont_non_analyt} and \eqref{eq:imYv}]. By contrast, if at least one of the two symmetries is present, then $\delta\mathrm{Im}\,Y_v \propto B - B_\mathrm{c}$ (see Appendix \ref{sec:app_imYv}) and thus $\delta\mathrm{Im}\,Y_v$ can be disregarded in the vicinity of the critical point in comparison with  $\delta\mathrm{Im}\,Y_i$.

Summarizing the estimates for $\delta\mathrm{Im}\,Y_i$ and $\delta\mathrm{Im}\,Y_v$, we conclude that near the critical point,
\begin{equation}\label{eq:crit}
    \delta \mathrm{Im}\,Y \propto (B - B_\mathrm{c}) \ln \bigl(B_\mathrm{c}/|B - B_\mathrm{c}|\bigr),
\end{equation}
regardless of the symmetry of the system.

To conclude, we note that at $T > 0$ the nonanalytic feature in Eq.~\eqref{eq:crit} is smeared. A detailed quantitative study of such thermal smearing is beyond the scope of the present work.

\section{Microscopic evaluation of the parameters of the low-energy theory\label{sec:micro}}

It is instructive to study how a concrete microscopic model of a topological junction fits the universal description of Sec.~\ref{sec:model}--\ref{sec:non_diss}. To do that, we consider a single-band semiconducting quantum wire with strong Rashba spin-orbit coupling. We assume that two adjoined sections of the wire are coated by $s$-wave superconducting shells so that a Josephson junction is formed between them. The wire is subjected to a parallel magnetic field. To start with, we focus on the case of a perfectly transparent junction at zero phase bias. Such a setup is described by the mean-field many-body Hamiltonian,
\begin{equation} \label{eq:mbody}
    H = \frac{1}{2}\int dx \, \Psi^\dagger(x) \hat{H} \mathrm \Psi(x),
\end{equation}
where $\Psi = \bigl(\psi_\uparrow,\,\psi_\downarrow,\,\psi^\dagger_\downarrow,\,-\psi^\dagger_\uparrow \bigr)^T$, $\psi_\sigma$ is the annihilation operator of electrons with spin $\sigma = \, \uparrow$ or $\downarrow$, and
\begin{equation}\label{eq:NW}
    \hat{H} = \Bigl(\frac{p^2}{2m} - \mu + v p \sigma_z \Bigr) \tau_z + \Delta(x) \tau_x - B \sigma_x.
\end{equation}
Here, $\sigma_{x,y,z}$ ($\tau_{x,y,z}$) are the Pauli matrices in spin (Nambu) space, $p = -i\partial_x$, $\mu$ is the chemical potenital, $m$ is the effective mass, $v$ is the spin-orbit coupling constant, $\Delta(x) = \Delta\,\Theta(|x| - \ell / 2)$ is the proximity-induced pairing potential (where $\ell$ is the length of the junction), and $B$ is the Zeeman energy. Within our model, we neglect the orbital effects of the magnetic field. For the Hamiltonian \eqref{eq:NW}, the topological transition occurs at a critical value of the Zeeman energy $B_\mathrm{c} = (\Delta^2 + \mu^2)^{1/2}$; $B < B_\mathrm{c}$ corresponds to the trivial phase and $B > B_\mathrm{c}$ corresponds to the topological phase. Below we focus on the vicinity of the transition, $|B - B_\mathrm{c}| \ll B_\mathrm{c}$. The \textit{microscopic} charge current operator $J(x)$ corresponding to the Hamiltonian \eqref{eq:mbody} is given by
\begin{align}\label{eq:current_micro_orig}
    J(x) = \frac{ie}{4m}\bigl[&\Psi^\dagger(x) \partial_x \Psi(x) - \bigl(\partial_x \Psi^\dagger(x)\bigr) \Psi(x)\bigr]\notag \\[0.25em]
    - \frac{ev}{2}&\Psi^\dagger(x) \sigma_z \Psi(x).
\end{align}

Hamiltonian \eqref{eq:NW} is symmetric under the mirror reflection ${\cal M}_x$ and the antiunitary symmetry $\cal{R}$. In terms of the microscopic model, these symmetries are represented~by
\begin{subequations}\label{eq:symms}
\begin{align}
    {\cal M}_x &= \exp(i \pi \sigma_x / 2)\,\mathcal{P}_x, \\
    \mathcal{R} &= \exp(i \pi \sigma_z / 2)\,\mathcal{T} \label{eq:symmsR},
\end{align}
\end{subequations}
where ${\cal P}_x$ is a parity operator ($\mathcal{P}_x \,x = -x$) and $\mathcal{T} = i\sigma_y \cal{K}$ is a time-reversal operator ($\cal{K}$ denotes complex conjugation). Symmetries \eqref{eq:symms} directly correspond to the previously introduced symmetries ${\cal M}_x$ and $\cal{R}$ of the low-energy theory [see Eqs.~\eqref{eq:mirror}, \eqref{eq:antiunitary}]. Notice that, in addition to  parity ${\cal P}_x$, the operator of mirror reflection features a spin-rotation around the $x$-axis. This rotation compensates for the sign change of the spin-orbit term under ${\cal P}_x$. In the same way, $\cal{R}$ features a spin-rotation around the spin-orbit axis, which undoes the flipping of the magnetic field under time-reversal $\cal{T}$. On a fundamental level, symmetry $\mathcal{R}$ can be identified with a combination $\mathcal{M}_z \mathcal{T}$, where $\mathcal{M}_z$ is the operator of a mirror reflection in the $z$-direction. In the simple single-band case considered here [Eq.~\eqref{eq:NW}], the action of $\mathcal{M}_z$ boils down to a spin-flip and hence $\mathcal{R}$ is given by Eq.~\eqref{eq:symmsR}. In more general cases, it is essential that the operation $\mathcal{M}_z$ also includes  $z\rightarrow -z$ (see Appendix \ref{sec:app_R} for details). We note that the microscopic current transforms as $J(x)\overset{\mathcal{M}_x}{\rightarrow} -J(-x)$ and $J(x)\overset{\mathcal{R}}{\rightarrow} -J(x)$, \textit{i.e.}, it is odd under both ${\cal M}_x$ and $\cal{R}$ at $x = 0$. 

The symmetries ${\cal M}_x$ and $\cal{R}$ might be broken by various perturbations of the Hamiltonian \eqref{eq:NW}. In particular, a finite phase difference $\varphi$ across the junction breaks both ${\cal M}_x$ and $\cal{R}$. The presence of a scattering potential at the junction (described by a term $\hat{V} = u(x) \tau_z$ in the single-particle Hamiltonian) breaks ${\cal M}_x$ if $u(x) \neq u(-x)$ but always leaves $\cal{R}$ intact. An example of a perturbation that breaks $\cal{R}$ but not ${\cal M}_x$ is a \textit{magnetic} scatterer at the junction which has an antisymmetric magnetization profile, $\hat{V} = b_z(x) \sigma_z$ with $b_z(-x) = -b_z(x)$ \footnote{We note that for a generic magnetization profile, $b_z(-x) \neq - b_z(x)$, the magnetic barrier breaks both ${\cal M}_x$ and $\cal{R}$.}.

To bring this microscopic example into the general framework of Sec.~\ref{sec:model}, we relate the phenomenological parameters $g$, $\alpha$, $\kappa_{ij}$ to the parameters of the Hamiltonian $\hat{H}$ and its perturbations $\hat{V}$. However, doing this in full generality is a tedious task. We make a number of approximations that simplify the problem. First, we assume that the spin-orbit coupling is strong, $mv^2 \gg \Delta, B$, and that the junction is short, $\ell \ll \xi \equiv v / \Delta$. These two approximations allow us to suppress the $x$-dependence of the absolute value of the pairing potential in Eq.~\eqref{eq:NW}. For simplicity, we also take $\mu = 0$. Finally, we suppose that all possible local perturbations, \textit{e.g.},~the scattering potential $u(x)$ or the magnetic scatterer $b_z(x)$, are smooth on the scale of the Fermi wavelength $\lambda_F \sim 1 / (m v)$. In this limit, only the modes close to $p = 0$ in momentum space are important  for the calculation of the parameters $g$, $\alpha$, $\kappa_{ij}$. Then, the Hamiltonian \eqref{eq:NW} can be linearized and expressed approximately as
\begin{equation}\label{eq:Hsimp}
    \hat{H} \approx  v p \, \sigma_z \tau_z + \Delta \tau_x - B \sigma_x.
\end{equation}

Next, it is convenient to express the electron field operators in terms of Majorana fields. To this end, we introduce
\begin{subequations}\label{eq:M}
\begin{align}
    \chi_R(x) &= i [\psi_\uparrow(x) - \psi_\uparrow^\dagger(x)] / \sqrt{2} , \\
    \chi_L(x) &= [\psi_\downarrow(x) + \psi_\downarrow^\dagger(x)] / \sqrt{2} , \\
    \eta_R(x) &=  [\psi_\uparrow(x) + \psi_\uparrow^\dagger(x)] / \sqrt{2} , \\
    \eta_L(x) &= i [\psi_\downarrow^\dagger(x) - \psi_\downarrow(x)] / \sqrt{2} .
\end{align}
\end{subequations}
Using these relations together with Eq.~\eqref{eq:Hsimp} we rewrite the many-body Hamiltonian~\eqref{eq:mbody} as
\begin{align}\label{eq:HM}
    H &\approx \ \frac{1}{2} \int dx \, \chi^T \bigl[-iv \zeta_z \partial_x  + (B - \Delta) \zeta_y\bigr]\chi \notag\\
    &\quad + \frac{1}{2}\int dx \, \eta^T \bigl[-iv \zeta_z \partial_x  + (B + \Delta) \zeta_y\bigr]\eta,
\end{align}
where $\chi = (\chi_R,\,\chi_L)^T$, $\eta = (\eta_R,\,\eta_L)^T$, and $\zeta_{x,y,z}$ are the Pauli matrices in $R/L$ space. Eq.~\eqref{eq:HM} identifies $\chi_{R/L}$ and $\eta_{R/L}$ as the low-energy and high-energy modes of the theory, respectively.

In terms of the Majorana fields, the microscopic charge current operator is given by
\begin{align}
    J(x) = &\frac{ie}{2m} \bigl[ \chi^T\!(x) \partial_x \chi(x) + \eta^T\!(x) \partial_x \eta(x)\bigr] \notag\\[0.25em]
    &- iev\,\eta^T\!(x) \zeta_z \chi(x).\label{eq:current_micro}
\end{align}
To obtain the low-energy current operator $I$ of Eq.~\eqref{eq:current}, we must project the microscopic current at the position of the junction, $J(x = 0)$, onto the low-energy subspace. In the pristine case in which both ${\cal M}_x$ and $\cal{R}$ are present, the Hamiltonian is block-diagonal in the $\chi$, $\eta$ representation [see Eq.~\eqref{eq:HM}], and the projection simply gives $I = (ie/2m) \chi^T\!(0)\partial_x\chi(0)$. Then, we find $g = 0$, $\alpha = 0$, and $\kappa = \kappa_0\mathbbm{1}$ with $\kappa_0 = 1 / 2m$, in accordance with Table~\ref{tab:class}.

We proceed by evaluating the low-energy parameters in the presence of symmetry-violating perturbations. First, we assume that a nonzero phase difference $\varphi$ is applied across the junction, breaking both ${\cal M}_x$ and $\cal{R}$. In this case, the pairing potential $\Delta$ should be replaced in Eq.~\eqref{eq:Hsimp} with $\Delta(x) = \Delta e^{-i(\varphi/2)\tau_z \,\mathrm{sgn}\,x}$. It is then convenient to perform a gauge transformation $\Psi(x)\rightarrow e^{-i(\varphi/4)\tau_z \,\mathrm{sgn}\,x}\Psi(x)$ which makes the pairing potential spatially uniform at the expense of producing a \textit{local} term $\hat{V} = - (v \varphi  / 2) \sigma_z \delta(x)$ \footnote{Another effect of the gauge transformation $\Psi(x)\rightarrow e^{-i(\varphi/4)\tau_z \,\mathrm{sgn}\,x}\Psi(x)$ is the appearance of a diamagnetic term in the current operator [which results from acting with $\partial_x$ in the first line of Eq.~\eqref{eq:current_micro_orig} on the exponent in the gauge transformation]. This term is negligible in comparison with the one in the second line of Eq.~\eqref{eq:current_micro_orig} provided $\ell \gg \lambda_F$ (where $\ell$ is the length of the junction and $\lambda_F$ is the Fermi wavelength) and $\varphi \lesssim 1$. We disregard the diamagnetic term in the following.}. Expressed in terms of the Majorana fields, the corresponding contribution to the many-body Hamiltonian is
\begin{equation}\label{eq:phase}
    V = i \frac{v \varphi}{2}\, \eta^T\!(0) \zeta_z \chi(0).
\end{equation}
Thus the phase bias directly couples low-energy and high-energy degrees of freedom. To project the Hamiltonian [defined by Eqs.~\eqref{eq:HM} and \eqref{eq:phase}] onto the low-energy subspace, we perform a Schrieffer-Wolff transformation $U_\varphi$ that removes this coupling to the first order in $\varphi \ll 1$ (see Appendix \ref{sec:app_phase}). This leads to a low-energy Hamiltonian of the form $H_\mathrm{w} + \delta H_\mathrm{w} + H_\mathrm{sc}$ [see Eqs.~\eqref{eq:Hw}, \eqref{eq:scat}, \eqref{eq:mass}] with $v_R = v_L = v$, $M = B - \Delta$, and\\
\begin{equation}\label{eq:g_phi}
    g \approx \frac{v\varphi^2}{8}.
\end{equation}
Notice that $g > 0$, so a finite $\varphi$ results in the presence of a discrete state at the junction in the trivial phase, $B < B_\mathrm{c} = \Delta$ [see Sec.~\ref{sec:away}]. The low-energy current is obtained by applying $U_\varphi$ to the microscopic current operator $J (x = 0)$. To lowest order in a gradient expansion we find $I \approx i e \alpha \, \chi_R(0)\chi_L(0)$ with $\alpha \approx -v\varphi / 2$. The presence of $\alpha \neq 0$ when ${\cal M}_x$ and $\cal{R}$ are absent is in accord with Table \ref{tab:class}. Another example of a perturbation that breaks both ${\cal M}_x$ and $\cal{R}$ is a component of the magnetic field along the spin-orbit axis; it is analyzed in detail in Appendix~\ref{sec:app_bz}.

Next, we analyze the low-energy theory in the presence of a scattering potential $\hat{V} = u(x)\tau_z$, \textit{i.e.}, the perturbation that generally breaks ${\cal M}_x$ but not $\cal{R}$. The many-body form of this perturbation in the Majorana basis is
\begin{equation}
    V = -i \int dx \, u(x) \, \eta^T\!(x) \chi(x).
\end{equation}
By decoupling $\chi$ and $\eta$ with a Schrieffer-Wolff transformation $U_u$ perturbatively in $u(x)$ (see Appendix \ref{sec:app_scat}) we find the parameters of the low-energy theory. First, 
\begin{equation}\label{eq:g_u}
    g \approx - \frac{1}{2v}\int \frac{d(q\xi)}{2\pi}\frac{4|u_q|^2}{4 + q^2\xi^2},
\end{equation}
where $\xi = v/\Delta$ and $u_q = \int dx \, u(x)e^{-iqx}$. Given $g < 0$, the scattering at the junction leads to the formation of a discrete state in the topological phase, $B > B_\mathrm{c}$ [see Sec.~\ref{sec:away}]. This is a shallow bound state described in~Ref.~\cite{murthy2020}. The low-energy current is given by Eq.~\eqref{eq:current} with $\alpha = 0$ and $\kappa = \kappa_0 \mathbbm{1} + \kappa_x\zeta_x$, where
\begin{align}\label{eq:kappa_0_u}
    \kappa_0 &\approx \frac{1}{2m} - \frac{v}{\Delta} \int \frac{d(q\xi)}{2\pi} \frac{u_q - q \partial_q u_q}{4 + q^2\xi^2},\\
    \kappa_x &\approx -\frac{v}{2\Delta} \int \frac{d(q\xi)}{2\pi} \frac{4i \xi^{-1} \partial_q u_q}{4+ q^2\xi^2}.\label{eq:kappa_x_u}
\end{align}
Notice that for a symmetric scattering potential [$u(x) = u(-x)$] the parameter $\kappa_x = 0$. This is a consequence of the restored mirror symmetry, cf.~Table \ref{tab:class}.

Finally, we consider a perturbation that breaks $\cal{R}$ but not ${\cal M}_x$: a magnetic barrier with an antisymmetric magnetization profile. In terms of the Majorana fields, it is described by
\begin{equation}
    V = -i\int dx \, b_z(x) \eta^T\!(x) \zeta_z \chi(x),
\end{equation}
where $b_z(-x) = -b_z(x)$. Projecting onto the low-energy subspace (see Appendix \ref{sec:app_mbarrier}), we obtain
\begin{equation}\label{eq:g_b_z}
   g \approx \frac{1}{2v} \int \frac{d(q\xi)}{2\pi} \frac{4|b_z^q|^2}{4 + q^2\xi^2},
\end{equation}
where $b_z^q = \int dx \, b_z(x)e^{-iqx}$. The low-energy current has the form \eqref{eq:current} with $\alpha = 0$ and $\kappa = \kappa_0 \mathbbm{1} + i\kappa_y \zeta_y$, where $\kappa_0 =  1 /2m$ and
\begin{equation}\label{eq:kappa_y_b_z}
    \kappa_y \approx  \frac{v}{2\Delta}\int \frac{d(q\xi)}{2\pi} \frac{4 i \xi^{-1} \partial_q b_z^q}{4 + q^2\xi^2}.
\end{equation}

\section{Discussion and Conclusions \label{sec:signatures}}
 
\begin{figure}[t]
  \begin{center}
    \includegraphics[scale = 1.0]{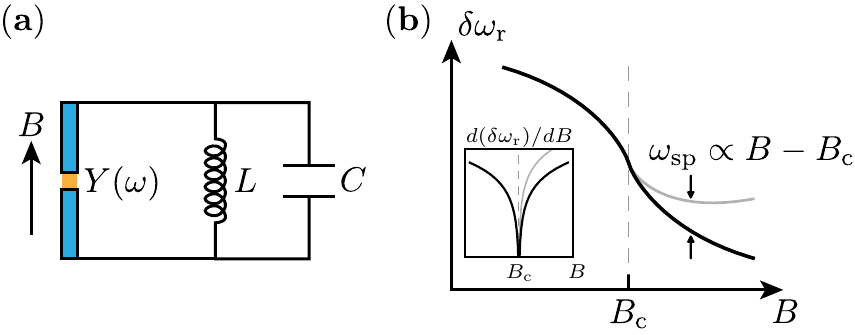}
    \caption{(a) Schematic drawing of a setup for probing the admittance of the junction in the cQED setting. The wire junction, subjected to the magnetic field $B$, is coupled to a microwave resonator [depicted as an $LC$-circuit]. Due to the coupling, the frequency of the resonator is shifted by $\delta \omega_\mathrm{r} \propto \mathrm{Im}\,Y(\omega_\mathrm{r})$. This type of measurement scheme at $B = 0$ was used in \cite{hays2018, hays2020}. (b) Sketch of the dependence of the resonator frequency shift $\delta \omega_\mathrm{r}$ on $B$ close to the critical point, $B = B_\mathrm{c}$. Solid black curve represents $\delta\omega_\mathrm{r}$ in the ground state of the junction. It has a nonanalytic feature, $\propto (B-B_\mathrm{c})\ln (B_\mathrm{c}/|B - B_\mathrm{c}|)$ [the nonanalyticity is highlighted in the inset, which shows the derivative of the frequency shift with respect to $B$]. The frequency shift in the poisoned state (solid gray curve) differs from the shift in the ground state, resulting in a fork-like feature; see Sec.~\ref{sec:signatures} for details.}
    \label{fig:resonator}
  \end{center}
\end{figure}

We have developed a theory of the dynamic electromagnetic response of a Josephson junction at the topological transition. We have found that the dissipative part of the junction's admittance may depend strongly on frequency and temperature, $\mathrm{Re}\,Y(\omega) \propto \max (\omega, T)^2$, despite the low-energy density of states approaching a constant at the critical point. This unusual critical behavior is observed if the system has at least one of two symmetries: the mirror symmetry ${\cal M}_x$ and the antiunitary symmetry ${\cal R}$. Only when ${\cal M}_x$ and ${\cal R}$ are both absent is the standard behavior---akin to the dissipative conductance of a normal metal---recovered. In this case, the dissipative part of the admittance is frequency- and temperature-independent, $\mathrm{Re}\,Y(\omega) = \mathrm{const}$. 

We have also extended our theory to study the electromagnetic response of the junction away from the critical point. We have established the scaling relations for the dependence of the admittance on frequency, temperature, and detuning of the magnetic field from its critical value, $B - B_\mathrm{c}$ [or, more generally, detuning of any control parameter from its critical value, see discussion around Eq.~\eqref{eq:mass}]. Interestingly, our approach predicts a pronounced asymmetry of the dissipative part of the admittance across the topological transition [see Fig.~\ref{fig:rey}]. The asymmetry results from the emergence of a non-degenerate in-gap state localized at the junction on one side of the transition.

The admittance of the topological junction can be readily accessed experimentally using the standard toolbox of circuit quantum electrodynamics (cQED). If the junction shunted by a high-$Q$ microwave resonator [see Fig.~\ref{fig:resonator}(a)], the resonant frequency of the latter, $\omega_\mathrm{r}$, is shifted  by an amount $\delta\omega_\mathrm{r}\propto \mathrm{Im}\,Y(\omega_\mathrm{r})$. Our theory predicts that the dependence of the frequency shift $\delta\omega_\mathrm{r}$ on the magnetic field $B$ should have a nonanalytic feature [see Eq.~\eqref{eq:crit}] at the critical point [see solid black line in Fig.~\ref{fig:resonator}(b)]. The observation of such a nonanalyticity may serve as an experimental signature of the topological transition. We note however that finite temperature of the device or finite length of the wire would tend to smooth out the sharp feature at $B = B_\mathrm{c}$.

We predict yet another effect which might be used to detect the topological transition: an incoherent splitting of the resonator frequency across $B = B_\mathrm{c}$ [see the fork-like feature in Fig.~\ref{fig:resonator}(b)]. The origin of such splitting can be explained in the following way. According to our theory, there is a bound state that is present at the junction on one side of the topological transition. When present, the bound state might occasionally trap a thermal or non-equilibrium quasiparticle \cite{hays2020}. The admittance of the junction in such a ``poisoned'' state (denoted $\widetilde{Y}$ below) differs from the admittance in a state with no trapped quasiparticles ($Y$). Accordingly, a single-shot measurement of the resonator frequency might yield two different outcomes with a difference $\omega_\mathrm{sp} \propto \mathrm{Im}\,\widetilde{Y}(\omega_\mathrm{r}) - \mathrm{Im}\,Y(\omega_\mathrm{r})$. Upon superimposing the results of many measurements, the difference results in an apparent splitting on one side of the topological transition. As we show in Appendix~\ref{sec:app_splitting}, in the vicinity of the transition, the splitting $\omega_\mathrm{sp} \propto \Theta(-M \cdot g)(B - B_\mathrm{c})$ \footnote{We note that in the topological phase the incoherent splitting of the resonator frequency might coexist with a coherent splitting due to the fermion parity mixing at the junction \cite{ginossar2014, keselman2019}.}.

The dissipative part of the admittance of the junction can also be accessed in the considered cQED architecture. When the junction is coupled to the resonator, the resonator's $Q$-factor is decreased by $\delta Q \propto \mathrm{Re}\,Y(\omega_\mathrm{r})$. Therefore scaling relations \eqref{eq:scY1}, \eqref{eq:scY2} for $\mathrm{Re}\,Y$ can be tested directly by measuring the $Q$-factor of the resonator with high precision. Our results indicate that the form of the scaling relations is extremely sensitive to the symmetries of the system. Thus it would be interesting to study in the laboratory how the dissipative response changes when the symmetries are broken in a controllable way, for example, by applying a phase bias across the junction or by changing the direction of the magnetic field with respect to the axis of the wire.

We expect that, on a qualitative level, our results for the critical behavior of the electromagnetic response (such as the suppressed dissipation in the presence of symmetries) may also be applicable to other experimentally relevant setups, \textit{e.g.}, a setup similar to that in Fig.~\ref{fig:resonator}(a) in which the Josephson junction is replaced by a uniform wire.

An interesting extension of our work may be to study the effects of disorder on the dynamic response of the topological junction. Disorder has a pronounced influence on the properties of one-dimensional superconducting wires with broken time-reversal and spin-rotation symmetries. In particular, it may induce a singularity in the density of states at the Fermi level if the wire is tuned sufficiently close to the topological transition \cite{motrunich2001, brouwer2011}.
It would be interesting to study how such a singularity, accompanied by the disorder-induced sample-to-sample fluctuations, affects the critical behavior of the junction's admittance.

All in all, we have revealed interesting and novel critical behavior in the electromagnetic response of a topological Josephson junction. Our theoretical results pave the way for the microwave detection of the topological phase transition in proximitized semiconducting nanowires, as an alternative to the identification of the topological phase via charge transport or thermodynamic signatures.

\acknowledgements{
We acknowledge very useful discussions with Valla Fatemi, Max Hays, Angela Kou, Roman Lutchyn, and Charlie Marcus. This work is supported by the DOE Contract No. DE-FG02-08ER46482 (LG), by the ARO grant W911NF-18-1-0212 (VK), by the Gordon and Betty Moore Foundation’s EPiQS Initiative through GBMF8686 (CM), and by the Microsoft Corporation.
}

\bibliography{references}

\clearpage
\appendix

\widetext
\section{Calculation of the dissipative part of the admittance in the vicinity of the critical point \label{sec:app_admittance}}

In this Appendix we present a detailed calculation of the dissipative part of the admittance within the framework of the low-energy theory. We begin by representing the Kubo formula for $\mathrm{Re}\,Y(\omega)$ [see Eqs.~\eqref{eq:kubo}, \eqref{eq:resp}]  in terms of the Green's functions of the helical Majorana modes:
\begin{equation}\label{eq:app_kubo}
    \mathrm{Re}\,Y(\omega) =  -\frac{2}{\pi\omega} \int_{-\infty}^{+\infty}dE \, \bigl[n(-E) - n(\omega - E)\bigr]\mathrm{Tr}\bigl\{\bigl[G^R(\omega - E) - G^A(\omega - E)\bigr]\hat{I} \bigl[G^R(-E) - G^A(- E)\bigr] \hat{I}\,\bigr\}.
\end{equation}
Here $n(E)$ is the quasiparticle distribution function. $G^{R/A}(E)$ is a retarded/advanced Green's function defined by
\begin{equation}
    G^{R/A}(E) = \frac{1}{E - \hat{H} \pm i0^+},
\end{equation}
where $0^+$ is an infinitesimally small positive number and
\begin{equation}\label{eq:app_singlep_h}
\hat{H} =   -i\hat{v}\partial_x + M \zeta_y + g\delta(x)\zeta_y, \quad 
\hat{v} = 
\begin{pmatrix}
v_R & 0 \\
0 & -v_L
\end{pmatrix},\quad \zeta_y =
\begin{pmatrix}
0 & -i \\
i & 0
\end{pmatrix}
\end{equation}
is a single-particle representation of the many-body Hamiltonian $H = H_\mathrm{w} + H_\mathrm{sc} + \delta H_\mathrm{w} $ [see Eqs.~\eqref{eq:Hw}, \eqref{eq:scat}, and \eqref{eq:mass}], in which matrices $\zeta_{y},\hat{v}$ act in the right-/left-mover subspace. The trace in Eq.~\eqref{eq:app_kubo} is taken over all single particle states (it includes the trace over the matrix indices). Finally, $\hat{I}$ is a single-particle representation of the low-energy current operator $I$ at $x = 0$ [in the coordinate representation, $\hat{I}(x)$ is defined in such a way that $I = \frac{1}{2}\int dx\,\chi_i(x) \hat{I}_{ij}(x)\chi_j(x)$]. The form of $\hat{I}$ is determined by the symmetries of the system. In case both ${\cal M}_x$ and $\cal{R}$ are absent, the gradient expansion of the current starts with a term with no derivatives and the single-particle current $\hat{I}$ is given by
\begin{equation}\label{eq:app_I_nosym}
    \hat{I}(x) \approx -e \alpha\,\zeta_y \delta(x)
\end{equation}
[see Eq.~\eqref{eq:current} and Table \ref{tab:class}]. Here, $e$ is the elementary charge, $\alpha$ is a proportionality coefficient that depends on the microscopic details of the system, and $\delta(x)$ is a Dirac delta-function. If at least one of the two symmetries ${\cal M}_x$ and $\cal{R}$ is present, $\alpha = 0$ and the gradient expansion of the low-energy current operator starts with derivative terms. In this case, $\hat{I}$ can be represented as
\begin{equation}\label{eq:app_I_sym}
    \hat{I}(x) = i e (\kappa_{0}\mathbbm{1}+\kappa_{x}\zeta_{x})\bigl[\delta(x)\overrightarrow{\partial_x}-\overleftarrow{\partial_x}\delta(x)\bigr] - e \kappa_{y}\zeta_{y}\bigl[\delta(x)\overrightarrow{\partial_x}+\overleftarrow{\partial_x}\delta(x)\bigr],
\end{equation}
where $\overrightarrow{\partial_x}$ and $\overleftarrow{\partial_x}$ are derivatives acting on the right and on the left, respectively, and $\mathbbm{1}$ is a unit matrix in the right-/left-mover subspace. We note that $\kappa_x = 0$ if ${\cal M}_x$ is a symmetry of the system and $\kappa_y = 0$ if $\cal{R}$ is a symmetry [see Table \ref{tab:class}].

We start by finding the exact expressions for the Green's functions [Sec.~\ref{sec:app_GF}]. Then, we use these expressions to compute the dissipative part of the admittance through Eq.~\eqref{eq:app_kubo} [Sec.~\ref{sec:app_adm}].

\subsection{The Green's function of Majorana modes\label{sec:app_GF}}

In the coordinate representation, the Green's function $G^{R/A}(x,x^\prime|E) = \langle x| G^{R/A}(E)|x^\prime \rangle$ is a $2 \times 2$ matrix (acting in the right-/left-mover subspace)
which satisfies the following Schr\"{o}dinger equation: 
\begin{equation}\label{eq:app_Schr}
    \left[E+i\hat{v}\partial_x - M \zeta_y - g\delta(x)\zeta_y\right]G^{R/A}(x,x^{\prime}|E) =\delta(x-x^{\prime})\mathbbm{1}
\end{equation}
[$G^{R}(x,x^\prime|E)$ and $G^{A}(x,x^\prime|E)$ are distinguished by their behavior at $x \rightarrow \pm \infty$, as discussed below]. The solution of Eq.~\eqref{eq:app_Schr} is qualitatively different for energies above the edge of the continuum, $|E| > E_\mathrm{gap}$, and for subgap energies, $|E| < E_\mathrm{gap}$ [$E_\mathrm{gap}$ is given by Eq.~\eqref{eq:gap}]. These two cases are considered separately below.

\subsubsection{Above-the-gap energies, $|E| > E_\mathrm{gap}$}

At energies above the continuum's edge, $|E| > E_\mathrm{gap}$, the Green's function is built of the plane-wave solutions of the homogeneous Schr\"{o}dinger equation $[-i\hat{v}\partial_x + M \zeta_y]\psi(x) = E \psi(x)$. We begin by finding these plane-wave solutions at $E > E_\mathrm{gap}$ which we then use to construct the Green's function at positive above-the-gap energies [the Green's function at $E < -E_\mathrm{gap}$ can be obtained with the help of particle-hole symmetry: $G^{R/A}_{\alpha\beta}(x, x^\prime|-E) = -[G^{R/A}_{\alpha\beta}(x, x^\prime|E)]^\star$].

By solving the homogeneous Schr\"{o}dinger equation, we find the right-moving [$\psi_R(x|E)$] and left-moving [$\psi_L(x|E)$] plane-wave states with energy $E > E_\mathrm{gap}$:
\begin{subequations}\label{eq:rl-wfs}
\begin{align}
\psi_R (x|E) &= \frac{\sqrt{2}}{\sqrt{v_R + v_L}}
\frac{e^{i(p_E + q_E) x}}{\sqrt{\mathrm{u}_E^2 - \mathrm{v}_E^2}} \left(\begin{array}{c}
\frac{\mathrm{u}_E}{\sqrt{1+r}}\\
\frac{i\mathrm{v}_E \,\mathrm{sgn}(M)}{\sqrt{1-r}}
\end{array}\right), \\
\psi_L(x|E) &= \frac{\sqrt{2}}{\sqrt{v_R + v_L}} 
\frac{e^{-i(p_E - q_E) x}}{\sqrt{\mathrm{u}_E^2 - \mathrm{v}_E^2}}
\left(\begin{array}{c}
-\frac{i\mathrm{v}_E \,\mathrm{sgn}(M)}{\sqrt{1+r}}\\
\frac{\mathrm{u}_E}{\sqrt{1-r}}
\end{array}\right) .
\end{align}
\end{subequations}
Here, the dimensionless number 
\begin{equation}\label{eq:app_r}
    r \equiv \frac{v_R - v_L}{v_R + v_L} \in [-1, 1]
\end{equation}
parameterizes the difference in velocities of helical modes ($r = 0$ when $v_R = v_L$), $\mathrm{u}_E, \mathrm{v}_E$ are Bogoliubov amplitudes,
\begin{subequations}\label{eq:app_bogoliubov}
\begin{align}
\mathrm{u}_E &= \frac{1}{\sqrt{2}} 
\Bigl(1 + \sqrt{1 - E_\mathrm{gap}^2/E^2}\Bigr)^{1/2}, \\
\mathrm{v}_E &= \frac{1}{\sqrt{2}} 
\Bigl(1 - \sqrt{1 - E_\mathrm{gap}^2/E^2}\Bigr)^{1/2},
\end{align}
\end{subequations}
and momenta $p_E$ and $q_E$ are given by
\begin{align}
p_E &= \frac{v_R + v_L}{2 v_R v_L} \sqrt{E^2 - E_\mathrm{gap}^2}, \\
q_E &= - \, \frac{v_R - v_L}{2v_R v_L} \, E.
\end{align}
The prefactors in Eq.~\eqref{eq:rl-wfs} are chosen in such a way that each plane wave solution carries unit probability current.

We now use the plane-wave solutions to obtain an expression for the retarded Green's function at $E > E_\mathrm{gap}$ [the advanced Green's function can be obtained from the retarded one through $G^A_{\alpha\beta} (x,x^\prime|E) = [G^R_{\beta\alpha}(x^\prime, x|E)]^\star$]. As follows from Eq.~\eqref{eq:app_Schr}, at all points $x$ (except for $x = 0$ and $x = x^\prime$) the Green's function can be represented as a linear combination of $\psi_R(x|E)$ and $\psi_L(x|E)$. The retarded character of the Green's function implies that only outgoing waves should be present at $x \rightarrow \pm \infty$. Thus, assuming at first $x^\prime > 0$, we can express $G^R(x,x^\prime|E)$ as:
\begin{equation}\label{eq:app_GF_pws}
   G^R_{\alpha\beta}(x,x^\prime|E) =
   \begin{cases}
    {\cal A}_\beta(x^\prime) \psi_R^\alpha(x|E),     &   x > x^\prime,\\
    {\cal B}_\beta(x^\prime) \psi_L^\alpha(x|E) + {\cal C}_\beta(x^\prime) \psi_R^\alpha(x|E),   &   0 < x < x^\prime, \\
    {\cal D}_\beta(x^\prime) \psi_L^\alpha(x|E),     &   x < 0,
   \end{cases}
\end{equation}
where $\alpha, \beta = \pm 1$ are spinor indexes corresponding to the value of $\zeta_z$. The eight coefficients ${\cal A}_\beta, {\cal B}_\beta, {\cal C}_\beta, {\cal D}_\beta$ can be found by matching the Green's function across $x = 0$ and $x = x^\prime$. To establish the matching conditions at $x = 0$, we regularize the scattering term in Eq.~\eqref{eq:app_Schr} as $g \delta(x) \to (g/2\epsilon) \,\Theta(\epsilon - \abs{x})$, where $\Theta(x)$ is the Heaviside step function, integrate both sides of the resulting equation over the interval $(-\epsilon, \epsilon)$, and take the limit $\epsilon \rightarrow 0$. This procedure yields
\begin{equation}\label{eq:app_match_0}
    G^{R}(0^+,x^{\prime}|E)={\cal T}G^{R}(0^-,x^{\prime}|E), \qquad 
    {\cal T} = \left(\begin{array}{cc}
    \cosh\bigl[\frac{g}{\sqrt{v_R v_L}}\bigr] & -\sqrt{\frac{1-r}{1+r}}\sinh\bigl[\frac{g}{\sqrt{v_R v_L}}\bigr]\\
    -\sqrt{\frac{1+r}{1-r}}\sinh\bigl[\frac{g}{\sqrt{v_R v_L}}\bigr] & \cosh\bigl[\frac{g}{\sqrt{v_R v_L}}\bigr]
    \end{array}\right),
\end{equation}
where $0^+$ ($0^-$) is an infinitesimally small positive (negative) number. To obtain the matching condition at $x = x^\prime$ we integrate both sides of Eq.~\eqref{eq:app_Schr} over the interval $x \in (x^\prime - \epsilon, x^\prime + \epsilon)$ and take $\epsilon \rightarrow 0$. This yields
\begin{equation}\label{eq:app_match_xprime}
    G^{R}(x^{\prime} + 0^+,x^{\prime}|E)-G^{R}(x^{\prime} - 0^+,x^{\prime}|E)=-i \hat{v}^{-1},
\end{equation}
where the matrix $\hat{v}$ was defined in Eq.~\eqref{eq:app_Schr}. Next, by enforcing matching conditions \eqref{eq:app_match_0} and \eqref{eq:app_match_xprime} upon the plane-wave decomposition \eqref{eq:app_GF_pws}, we find the expression for the Green's function at $E > E_\mathrm{gap}$ and $x^\prime > 0$:
\begin{equation} \label{eq:app_gf_p_e}
    G^{R}(x,x^{\prime}|E) = -\frac{2i}{v_R + v_L}\frac{1}{\mathrm{u}_E^2 - \mathrm{v}_E^2}
    \begin{cases}
        \hat{\Pi}^{+-}_E e^{i(p_{E}+q_{E})(x-x^{\prime})}-\hat{\Pi}^{++}_E e^{ip_{E}(x+x^{\prime})}e^{iq_{E}(x-x^{\prime})}{\cal V}_{E}\sinh\bigl[\frac{g}{\sqrt{v_R v_L}}\bigr], & x > x^\prime,\\
        \hat{\Pi}^{-+}_E e^{i(q_{E}-p_{E})(x-x^{\prime})}-\hat{\Pi}^{++}_E e^{ip_{E}(x+x^{\prime})}e^{iq_{E}(x-x^{\prime})} {\cal V}_{E}\sinh\bigl[\frac{g}{\sqrt{v_R v_L}}\bigr], & 0 < x < x^{\prime},\\
       \hat{\Pi}^{-+}_E e^{i(q_{E}-p_{E})(x-x^{\prime})}{\cal V}_{E}(\mathrm{u}_E^2-\mathrm{v}_E^2), & x < 0,\\
    \end{cases}
\end{equation}
where we introduced matrices
\begin{subequations}
\begin{align}
\hat{\Pi}^{++}_E &= \left(\begin{array}{cc}
\frac{i\mathrm{u}_E\mathrm{v}_E \,\mathrm{sgn}(M)}{1+r} & \frac{\mathrm{u}_E^{2}}{\sqrt{1-r^{2}}}\\
-\frac{\mathrm{v}_E^{2}}{\sqrt{1-r^{2}}} & \frac{i\mathrm{u}_E\mathrm{v}_E \,\mathrm{sgn}(M)}{1-r}
\end{array}\right), \\
\hat{\Pi}^{+-}_E &= \left(\begin{array}{cc}
\frac{\mathrm{u}_E^{2}}{1+r} & -\frac{i \mathrm{u}_E\mathrm{v}_E \,\mathrm{sgn}(M)}{\sqrt{1-r^{2}}}\\
\frac{i \mathrm{u}_E\mathrm{v}_E \,\mathrm{sgn}(M)}{\sqrt{1-r^{2}}} & \frac{\mathrm{v}_E^{2}}{1-r}
\end{array}\right),
\end{align}
\end{subequations}
and $\hat{\Pi}^{-+}_E$, which is obtained from $\hat{\Pi}^{+-}_E$ by the interchange of $\mathrm{u}_E$ and $\mathrm{v}_E$. The function ${\cal V}_E$ is given by \begin{equation}
    {\cal V}_E = \frac{1}{(\mathrm{u}_E^2 - \mathrm{v}_E^2)\cosh \bigl[\frac{g}{\sqrt{v_R v_L}}\bigr] + 2i\,\mathrm{sgn}(M)\mathrm{u}_E \mathrm{v}_E \sinh \bigl[\frac{g}{\sqrt{v_R v_L}}\bigr]}.
\end{equation}
Similarly, for $x^\prime < 0$ we obtain
\begin{equation}\label{eq:eq:app_gf_p_e_2}
    G^{R}(x,x^{\prime}|E) = -\frac{2i}{v_R + v_L}\frac{1}{\mathrm{u}_E^2 - \mathrm{v}_E^2}
    \begin{cases}
        \hat{\Pi}^{+-}_E e^{i(p_{E}+q_{E})(x-x^{\prime})}{\cal V}_{E}(\mathrm{u}_E^2-\mathrm{v}_E^2), & x > 0,\\
        \hat{\Pi}^{+-}_E e^{i(p_{E} + q_{E})(x-x^{\prime})}-\hat{\Pi}^{--}_E e^{-ip_{E}(x+x^{\prime})}e^{iq_{E}(x-x^{\prime})} {\cal V}_{E}\sinh\bigl[\frac{g}{\sqrt{v_R v_L}}\bigr], & x^\prime < x < 0,\\
       \hat{\Pi}^{-+}_E e^{i(q_{E} - p_{E})(x-x^{\prime})}-\hat{\Pi}^{--}_E e^{-ip_{E}(x+x^{\prime})}e^{iq_{E}(x-x^{\prime})}{\cal V}_{E}\sinh\bigl[\frac{g}{\sqrt{v_R v_L}}\bigr], & x < x^\prime,\\
    \end{cases}
\end{equation}
where $\hat{\Pi}^{--}_E$ is obtained from $\hat{\Pi}^{++}_E$ by the interchange of $\mathrm{u}_E$ and $\mathrm{v}_E$.

\subsubsection{Subgap energies, $|E| < E_\mathrm{gap}$\label{sec:app_ingap}}

To find the Green's function at $|E| < E_\mathrm{gap}$, we first establish the structure of the subgap spectrum by directly solving the Schr\"{o}dinger equation:
\begin{equation}
    [-i\hat{v}\partial_x + M \zeta_y + g\delta(x) \zeta_y]\psi(x) = E\psi(x).
\end{equation}
If present, a subgap solution should have the following structure:
\begin{equation}
    \psi(x) =
    \begin{cases}
     {\cal A} \psi_D(x|E),     & x > 0,\\
     {\cal B} \psi_G(x|E),     & x < 0,
    \end{cases}
\end{equation}
where
\begin{equation}
    \psi_{D/G}(x|E) = e^{\mp x\frac{v_R + v_L}{2v_R v_L}
    \sqrt{E_\mathrm{gap}^{2}-E^{2}}} e^{iq_{E}x}
    \left(\begin{array}{c}
    \frac{e^{\mp i\theta (E)}}{\sqrt{1+r}}\\
    \frac{i\,\mathrm{sgn}(M)}{\sqrt{1-r}}
    \end{array}\right),
\end{equation}
${\cal A}$ and ${\cal B}$ are coefficients that are yet to be determined, and
\begin{equation}
    e^{i\theta (E)} = E / E_\mathrm{gap} - i\sqrt{1 - E^2 / E_\mathrm{gap}^2}.
\end{equation}
The wavefunction $\psi(x)$ should be matched across $x = 0$ through the relation $\psi(0+) = {\cal T} \psi(0-)$, where the transfer matrix ${\cal T}$ is defined in Eq.~\eqref{eq:app_match_0}. The matching results in a system of linear equations for ${\cal A}$ and ${\cal B}$ that has a solution only if the energy satisfies
\begin{equation}\label{eq:app_bs_eq}
    \sin (\theta (E)) = \mathrm{sgn}(M) \tanh\Bigl[ \frac{g}{\sqrt{v_R v_L}}\Bigr].
\end{equation}
If $M \cdot g > 0$ the right hand side and the left hand side have different signs so the equality is never satisfied at $|E| < E_\mathrm{gap}$. Consequently, there is no bound state at the junction. By contrast, if $M \cdot g < 0$, Eq.~\eqref{eq:app_bs_eq} has a non-degenerate solution $E = \pm E_\tau$, where $E_\tau = E_\mathrm{gap} \sqrt{\tau}$ and $\tau = 1/\cosh^2[g/\sqrt{v_R v_L}]$ [see Eq.~\eqref{eq:bound}]. The wavefunction $\psi_\tau (x|E_\tau)$ at $E = E_\tau$ is given by
\begin{equation}\label{eq:app_bs_wf} 
\psi_\tau(x|E_\tau)	=\left[\frac{E_\mathrm{gap}\sqrt{1-\tau}}{v_R + v_L}\right]^{1/2}
e^{- \frac{v_R + v_L}{2v_R v_L} E_\mathrm{gap} \sqrt{1 - \tau} |x| + iq_{E_\tau} x}
\begin{cases}
\left(\begin{array}{c}
\frac{1}{\sqrt{1+r}}\\
\frac{ie^{i\theta (E_\tau)}\,\mathrm{sgn}(M)}{\sqrt{1-r}}
\end{array}\right) & x>0,\\
\left(\begin{array}{c}
\frac{e^{i\theta (E_\tau)}}{\sqrt{1+r}}\\
\frac{i\,\mathrm{sgn}(M)}{\sqrt{1-r}}
\end{array}\right) & x<0.
\end{cases}
\end{equation}
The wavefunction $\psi_\tau(x|-E_\tau)$ at $E = -E_\tau$  can be obtained from $\psi_\tau(x|E_\tau)$ with the help of the particle-hole symmetry. The latter acts as complex conjugation in the Majorana basis and thus   $\psi_\tau(x|-E_\tau) = \psi^\star_\tau(x|E_\tau)$.

To find the Green's function at subgap energies it is most convenient to employ its spectral representation. The retarded/advanced Green's function is given by
\begin{equation}
    G^{R/A}(x,x^\prime|E) = \sum_{n} \frac{\psi_n(x)\psi^\dagger_n(x^\prime)}{E - E_n \pm i0^+},
\end{equation}
where the sum is carried over all eigenstates of the single-particle Hamiltonian $\hat{H}$. Using the Sokhotski–Plemelj theorem, $G^{R/A}(x,x^\prime|E)$ can be represented as 
\begin{equation}
    G^{R/A}(x,x^\prime|E) = G^{(1)}(x,x^\prime|E) + G^{(2),R/A}(x,x^\prime|E),
\end{equation}
where
\begin{subequations}
\begin{align}
    G^{(1)}(x,x^\prime|E) &= \mathrm{P} \sum_n \frac{\psi_n(x)\psi^\dagger_n(x^\prime)}{E - E_n}, \\
    G^{(2),R/A}(x,x^\prime|E) &= \mp i\pi \sum_n \delta(E - E_n) \psi_n(x)\psi^\dagger_n(x^\prime),
\end{align}
\end{subequations}
and $\mathrm{P}$ denotes the Cauchy principal part. We note that the contribution $G^{(1)}(x,x^\prime|E)$ is the same for the retarded and advanced Green's functions and, therefore, cancels in the calculation of the dissipative part of the admittance [see Eq.~\eqref{eq:app_kubo}]. Thus we refrain from finding it explicitly here.

The structure of $G^{(2),R/A}(x,x^\prime|E)$ at subgap energies depends on the sign of  $M \cdot g$. If $M\cdot g > 0$ this contribution vanishes: there are no states below the gap and thus $\delta(E - E_n) = 0$ for all $n$ if $|E| < E_\mathrm{gap}$. By contrast, if $M\cdot g < 0$ there is a bound state at the junction which contributes to the sum in the definition of $G^{(2),R/A}(x,x^\prime|E)$. In this case, using Eq.~\eqref{eq:app_bs_wf} we find for $0 < E < E_\mathrm{gap}$:
\begin{gather}\label{eq:app_gf_bs}
    G^{(2),R/A}(x,x^\prime|E) = \mp i\pi  \frac{E_\mathrm{gap} \sqrt{1 - \tau}}{v_R + v_L}  e^{iq_{E_\tau} (x - x^\prime)} e^{-\frac{v_R + v_L}{2 v_R v_L} E_\mathrm{gap}\sqrt{1 - \tau}  (|x| + |x^\prime|)}\delta(E - E_\tau) \hat{Q}_{\mathrm{sgn}\,x, \mathrm{sgn}\,x^\prime},
\end{gather}
where the matrices $\hat{Q}_{ss^\prime}$ ($s,s^\prime = \pm$) are given by
\begin{subequations}
\begin{align}
    \hat{Q}_{++} &= \begin{pmatrix}
    \frac{1}{1+r} & \frac{-ie^{-i\theta (E_\mathrm{\tau})}\,\mathrm{sgn}(M)}{\sqrt{1-r^2}}\\
    \frac{ie^{i\theta (E_\mathrm{\tau})}\,\mathrm{sgn}(M)}{\sqrt{1-r^2}} & \frac{1}{1 - r}
    \end{pmatrix}, \\
    \hat{Q}_{+-} &= \begin{pmatrix}
    \frac{e^{-i\theta (E_\mathrm{\tau})}}{1+r} & \frac{-i\,\mathrm{sgn}(M)}{\sqrt{1-r^2}}\\
    \frac{i\,\mathrm{sgn}(M)}{\sqrt{1-r^2}} & \frac{e^{i\theta (E_\mathrm{\tau})}}{1 - r}
    \end{pmatrix},
\end{align}
\end{subequations}
and $\hat{Q}_{-+} = \zeta_z \hat{Q}_{+-}^\star \zeta_z$, $\hat{Q}_{--} = \zeta_z \hat{Q}_{++}^\star \zeta_z$. An expression for $G^{(2),R/A}$ at negative energies ($-E_\mathrm{gap} < E < 0$)  can be obtained from Eq.~\eqref{eq:app_gf_bs} with the help of the particle-hole symmetry: $G_{\alpha\beta}^{(2),R/A}(x,x^\prime|E) = -[G_{\alpha\beta}^{(2),R/A}(x,x^\prime|-E)]^\star$. 

\subsubsection{Local density of states\label{sec:app_dos}}

The expressions for the Green's functions [see Eqs.~\eqref{eq:app_gf_p_e}, \eqref{eq:app_gf_bs}] can be used to find the local density of states (DOS) at the position of the junction. The latter quantity is given by
\begin{equation}
    \nu_\mathrm{loc}(E) = \frac{i}{2\pi}\mathrm{tr}\bigl[G^R(0^+,0^+|E) - G^A(0^+,0^+|E) \bigr],
\end{equation}
where $\mathrm{tr}$ denotes the matrix trace. Notice that the Green's functions are evaluated at a point slightly displaced from $x,x^\prime = 0$ to the right---this is to avoid an ambiguity related to the discontinuity of $G^{R/A}(x,x^\prime|E)$ at these points (changing $0^+$ to $0^-$ does not change the result for the local DOS). Assuming $E > 0$, we find
\begin{equation}\label{eq:app_local_dos}
    \nu_\mathrm{loc}(E) = \Bigl[\frac{1}{2\pi v_R} + \frac{1}{2\pi v_L}\Bigr]\bigl(\pi \sqrt{1-\tau}E_\mathrm{gap}\Theta(-M\cdot g)\delta(E - E_\mathrm{\tau})  + \Theta(E - E_\mathrm{gap})\,\rho (E/E_\mathrm{gap})\bigr).
\end{equation}
The first term in the round brackets describes the contribution to the local DOS due to the bound state. It is present when $M \cdot g < 0$ only, as highlighted by the Heaviside step function. The second term describes the contribution to the local DOS due to states of the continuous part of the spectrum; $\rho(\varepsilon)$ is given by Eq.~\eqref{eq:ldos}.

\subsection{Calculation of the dissipative part of the admittance\label{sec:app_adm}}

The dissipative part of the admittance can now be computed by substituting the low-energy current operator [either Eq.~\eqref{eq:app_I_nosym} or Eq.~\eqref{eq:app_I_sym} depending on the symmetry of the system] and explicit expressions for the Green's functions [Eqs.~\eqref{eq:app_gf_p_e}, \eqref{eq:eq:app_gf_p_e_2},  \eqref{eq:app_gf_bs}] into the Kubo formula [Eq.~\eqref{eq:app_kubo}]. The only subtlety in the calculation is that the Green's functions are discontinuous at $x,x^\prime = 0$, \textit{i.e.}, at the point where the current operator is evaluated. To get a well-defined result for $\mathrm{Re}\,Y(\omega)$ we replace $\delta(x)$ in Eqs.~\eqref{eq:app_I_nosym}, \eqref{eq:app_I_sym} by $[\delta(x + 0^+) + \delta(x - 0^+)]/2$ \footnote{Other regularizations for the current operator are possible. It can be shown that the scaling functions do not depend on the choice of regularization, as long as the latter is consistent with the symmetries of the system.}; the rest of the calculation is straightforward. The resulting expression for $\mathrm{Re}\,Y(\omega)$ can be conveniently divided into four contributions corresponding to four types of energy absorption processes [see Fig.~\ref{fig:processes}]:

\begin{figure}[t]
  \begin{center}
    \includegraphics[scale = 1.0]{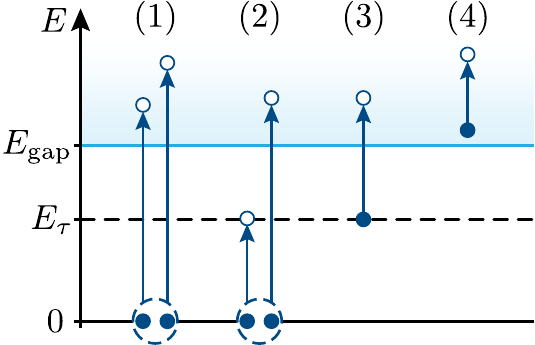}
    \caption{Four types of processes that contribute to $\mathrm{Re}\,Y(\omega)$. In processes of type $(1)$ a pair of quasiparticles is produced above the continuum's edge. In processes of type $(2)$ a pair of quasiparticles is produced, one at the in-gap state and one  above the continuum's edge. In processes of type (3) a quasiparticle is promoted from the in-gap state to the continuum. In processes of type (4) an energy quantum is absorbed by a quasiparticle in the continuum. The respective contributions to the dissipative part of the admittance are denoted $\mathrm{Re}\,Y^{(i)}(\omega)$ ($i=1,...,4$).}
    \label{fig:processes}
  \end{center}
\end{figure}

\begin{itemize}
    \item $\mathrm{Re}\,Y^{(1)}(\omega)$ corresponds to processes in which a Cooper pair in the condensate is broken by a drive photon into two quasiparticles at energies above the continuum's edge. On a formal level, this contribution originates from a part of the energy integral in Eq.~\eqref{eq:app_kubo} in which $E,\,\omega - E > E_\mathrm{gap}$. $\mathrm{Re}\,Y^{(1)}(\omega)$ is nonzero only at frequencies $\omega > \omega_\mathrm{th}^{(1)} = 2E_\mathrm{gap}$. 
    \item $\mathrm{Re}\,Y^{(2)}(\omega)$ is present when there is a bound state at the junction [which requires $M\cdot g < 0$; see discussion after Eq.~\eqref{eq:app_bs_eq}]. It describes processes of energy absorption in which a Cooper pair is broken into one quasiparticle above the continuum's edge and one quasiparticle at the bound state.  $\mathrm{Re}\,Y^{(2)}(\omega)$ corresponds to a part of the energy integral in Eq.~\eqref{eq:app_kubo} in which either $E > E_\mathrm{gap}$, $0 < \omega - E < E_\mathrm{gap}$ or $0 < E < E_\mathrm{gap}$, $\omega - E > E_\mathrm{gap}$. This contribution is nonzero only at $\omega > \omega_\mathrm{th}^{(2)} = E_\mathrm{gap} + E_\tau$, where $E_\tau$ is the energy of the bound state [see Eq.~\eqref{eq:bound}].
    \item $\mathrm{Re}\,Y^{(3)}(\omega)$ describes processes in which a quasiparticle at the bound state absorbs an energy quantum and gets promoted to the continuum [these processes also require $M\cdot g < 0$ to occur]. It originates from a part of the integral in which either $E > E_\mathrm{gap}$, $0< E - \omega  < E_\mathrm{gap}$ or $-E_\mathrm{gap} < E < 0$, $ E - \omega < - E_\mathrm{gap}$. The threshold frequency for this contribution is $\omega_\mathrm{th}^{(3)} = E_\mathrm{gap} - E_\tau$.
    \item Finally, $\mathrm{Re}\,Y^{(4)}(\omega)$ describes processes in which a quasiparticle in the continuum absorbs an energy quantum. The corresponding integration domain is defined by $E, E - \omega > E_\mathrm{gap}$ or $E - \omega, E < - E_\mathrm{gap}$. $\mathrm{Re}\,Y^{(4)}(\omega)$ does not have a frequency threshold: it is present at all $\omega$ provided there are quasiparticles above the continuum's edge.  
\end{itemize}
Note that all of the above processes involve the states of the continuum; there is no discrete line in the absorption spectrum associated with the in-gap state. The reason for the absence of a discrete line is that the non-degenerate in-gap state can accommodate only one quasiparticle, whereas in each relevant energy absorption event two quasiparticles are produced [there are also processes in which preexisting quasiparticles are excited, but such processes cannot lead to a discrete line in the absorption spectrum either, because there is only one in-gap state].

Under the assumption that the quasiparticle distribution function corresponds to thermal equilibrium at temperature $T$, it is possible to represent the four contributions to the dissipative part of the admittance in the scaling form:
\begin{subequations}\label{eq:app_contributions}
\begin{align}
    \mathrm{Re}\,Y^{(1)}(\omega) &= C \omega^\gamma f^{(1)}\Bigl(\frac{\omega}{E_\mathrm{gap}} , \frac{T}{E_\mathrm{gap}}\Bigr), \\
    \mathrm{Re}\,Y^{(2,3,4)}(\omega) &= C E_\mathrm{gap}\omega^{\gamma - 1} f^{(2,3,4)}\Bigl(\frac{\omega}{E_\mathrm{gap}}, \frac{T}{E_\mathrm{gap}}\Bigr).
\end{align}
\end{subequations}
Here $\gamma$ is the dynamic critical exponent [see Eqs.~\eqref{eq:dynexp}; $\gamma = 0$ in the absence of symmetries and $\gamma = 2$ if at least one of the two symmetries ${\cal M}_x$ and $\cal{R}$ is present]. $C$ is a constant factor that is determined by the coefficients in the gradient expansion of the current operator. In the absence of symmetries we find
\begin{equation}
    C = c_\alpha = G_0 \frac{\alpha^2 \tau}{4v_R v_L},
\end{equation}
where $\alpha$ is defined in Eq.~\eqref{eq:app_I_nosym} and $G_0 = e^2/\pi$ is the conductance quantum. In the presence of symmetries we obtain
\begin{equation}\label{eq:app_cs}
    C = 
    \begin{cases}
    c_0,\quad & \text{if ${\cal M}_x$ and $\cal{R}$ are both present,}\\
    c_0 + c_y,\quad & \text{if ${\cal M}_x$ is present but $\cal{R}$ is absent,}\\
    c_0 + c_x,\quad & \text{if $\cal{R}$ is present but ${\cal M}_x$ is absent,}
    \end{cases}
\end{equation}
where 
\begin{equation}
        c_0 = G_0\frac{\kappa_0^2\tau}{12v^4}, \quad c_x = G_0\frac{\kappa_x^2\tau}{4v^4}, \quad c_y = G_0\frac{\kappa_y^2}{12v^4}
\end{equation}
with parameters $\kappa_{0,x,y}$ defined in Eq.~\eqref{eq:app_I_sym} and $v = v_R = v_L$ [recall that the velocities of right- and left-moving modes are the same in the presence of symmetries; see Table \ref{tab:class}]. Finally, as follows from a direct calculation, the dimensionless scaling functions are given by
\begin{align}
    f^{(1)}(w, \mathrm{t}) &= \frac{\Theta(w - 2)}{w}\int_{1}^{w - 1} d\varepsilon \, \rho(\varepsilon)\rho(w - \varepsilon) |z(w, \varepsilon)|^2 \Bigl[1 - \tilde{n}\Bigl(\frac{w - \varepsilon}{\mathrm{t}}\Bigr) - \tilde{n} \Bigl(\frac{\varepsilon}{\mathrm{t}}\Bigr)\Bigr],\label{eq:app_scaling_1}\\
    f^{(2)}(w, \mathrm{t}) &=2\pi\Theta(-M\cdot g)\Theta(w - 1 - \sqrt{\tau}) \sqrt{1-\tau}\rho(w - \sqrt{\tau})|z(w, \sqrt{\tau})|^2 \Bigl[1 - p_\tau - \tilde{n}\Bigl(\frac{w - \sqrt{\tau}}{\mathrm{t}}\Bigr)\Bigr],\label{eq:app_scaling_2}\\
    f^{(3)}(w, \mathrm{t}) &=2\pi\Theta(-M\cdot g)\Theta(w - 1 + \sqrt{\tau})\sqrt{1 - \tau}\rho(w + \sqrt{\tau})|z(w, -\sqrt{\tau})|^2 \Bigl[p_\tau - \tilde{n}\Bigl(\frac{w + \sqrt{\tau}}{\mathrm{t}}\Bigr)\Bigr],\label{eq:app_scaling_3}\\
    f^{(4)}(w, \mathrm{t}) &= 2\int_1^{+\infty} d\varepsilon \, \rho(\varepsilon)\rho(\varepsilon + w) |z(w, -\varepsilon)|^2 \Bigl[\tilde{n}\Bigl(\frac{\varepsilon}{\mathrm{t}}\Bigr) - \tilde{n}\Bigl(\frac{\varepsilon + w}{\mathrm{t}}\Bigr)\Bigr].\label{eq:app_scaling_4}
\end{align}
In these expressions, $w = \omega / E_\mathrm{gap}$ is the dimensionless frequency of the drive and $\mathrm{t} = T / E_\mathrm{gap}$ is the dimensionless temperature.  The transition matrix element $|z(w,\varepsilon)|$ is controlled by the symmetry of the system: it is given by Eq.~\eqref{eq:matrixsymI} if ${\cal M}_x$ is present, by Eq.~\eqref{eq:matrixsymnoI} if ${\cal M}_x$ is absent but $\cal{R}$ is present, and by Eq.~\eqref{eq:matrixnosyms} if ${\cal M}_x$ and $\cal{R}$ are both absent. $\rho(\varepsilon)$ is the dimensionless local density of states above the continuum's edge [see Eq.~\eqref{eq:ldos} and Appendix \ref{sec:app_dos}]. $\tilde{n}(\varepsilon / \mathrm{t})$ is the Fermi-Dirac distribution function expressed in terms of the dimensionless variables,
\begin{equation}
    \tilde{n}\Bigl(\frac{\varepsilon}{\mathrm{t}}\Bigr) = \frac{1}{e^{\varepsilon / \mathrm{t}} + 1}.
\end{equation}
$p_\tau = \tilde{n}(\sqrt{\tau} / \mathrm{t})$ is the occupation probability of the bound state. The frequency-dependent step functions in Eqs.~\eqref{eq:app_scaling_1}--\eqref{eq:app_scaling_3} indicate that the contributions $\mathrm{Re}\,Y^{(1)}(\omega)$, $\mathrm{Re}\,Y^{(2)}(\omega)$, and $\mathrm{Re}\,Y^{(3)}(\omega)$ are nonzero only above the respective threshold frequencies. The factor $\Theta(-M\cdot g)$ in Eqs.~\eqref{eq:app_scaling_2} and \eqref{eq:app_scaling_3} highlights that the contributions $\mathrm{Re}\,Y^{(2)}(\omega)$ and $\mathrm{Re}\,Y^{(3)}(\omega)$ are only present when there is a bound state at the junction. The factor $\sqrt{1-\tau}$ in $f^{(2,3)}(w)$ originates from a subgap contribution to the density of states due to the bound state; see Eq.~\eqref{eq:app_local_dos}. The scaling functions $f^{(i)}(w, \mathrm{t})$ are factored out in Eq.~\eqref{eq:app_contributions} in such a way that a finite limit $f^{(i)}(w \rightarrow \infty, \mathrm{t})$ exists.

At zero temperature, $\mathrm{Re}\,Y^{(3,4)}(\omega) = 0$ because there are no thermally excited quasiparticles in the system to absorb the energy of a drive photon. By defining $f^{(1,2)}(w) \equiv f^{(1,2)}(w, \mathrm{t} = 0)$ from Eqs.~\eqref{eq:app_contributions}, \eqref{eq:app_scaling_1}, \eqref{eq:app_scaling_2} we arrive at Eqs.~\eqref{eq:scY1}, \eqref{eq:scaling}, \eqref{eq:scY2}, and \eqref{eq:scaling2} of the main text.

Equations \eqref{eq:app_contributions}, \eqref{eq:app_scaling_1}--\eqref{eq:app_scaling_4} can also be applied to find $\mathrm{Re}\,Y(\omega)$ at the critical point (at $T \neq 0$). To do that,  we take the limit  $E_\mathrm{gap} \rightarrow 0$ in Eq.~\eqref{eq:app_contributions}. In this limit, contributions $\mathrm{Re}\,Y^{(2)}(\omega)$ and $\mathrm{Re}\,Y^{(3)}(\omega)$ vanish. The two remaining contributions, $\mathrm{Re}\,Y^{(1)}(\omega)$ and $\mathrm{Re}\,Y^{(4)}(\omega)$, can be combined into a single integral over energy that can be easily performed for any symmetry of the system. The results of this calculation are presented in Eqs.~\eqref{eq:const}, \eqref{eq:critadm_0}--\eqref{eq:critadm_x} of the main text.

Finally, we note that Eqs.~\eqref{eq:app_contributions}, \eqref{eq:app_scaling_1}--\eqref{eq:app_scaling_4} can be straightforwardly generalized to describe the dissipative response of the junction in cases where the quasiparticle distribution function $n(E)$ does not correspond to thermal equilibrium (\textit{e.g.}, due to the presence of non-equilibrium quasiparticles in the device). This is achieved by replacing the Fermi-Dirac distribution function $\tilde{n}(\varepsilon / \mathrm{t})$ in Eqs.~\eqref{eq:app_scaling_1}--\eqref{eq:app_scaling_4} by the non-equilibrium distribution function $n(E_\mathrm{gap} \varepsilon)$.

\section{Evaluation of the non-dissipative part of the admittance}

In this Appendix, we present derivations of the results on the non-dissipative part of the admittance discussed in Secs.~\ref{sec:non_diss} and \ref{sec:signatures}.

\subsection{Derivation of a general expression for $\mathrm{Im}\,Y(\omega)$ \label{sec:app_linresp}}

We start by obtaining a general expression for the non-dissipative part of the admittance of a Josephson junction, which we used to get Eqs.~\eqref{eq:im_adm}--\eqref{eq:im_adm_v}. Let us denote the microscopic Hamiltonian of the junction by $H(\varphi)$, where $\varphi$ is the phase difference between the superconducting leads. The particular form of $H(\varphi)$ is not important for our derivation and is not specified below. For convenience, we assume that the gauge is fixed in such a way that the phase bias is described by a local term at the position of the junction, while the pairing potential in the leads is real. In this gauge, the alternating voltage $V(t)$ applied across the junction can be accounted for by taking $H(\varphi) \rightarrow H(\varphi + \delta \varphi(t))$, where $\delta\varphi(t)$ is related to $V(t)$ through the Josephson relation $\delta \dot{\varphi}(t) = 2e V(t)$ (here $e > 0$ is the elementary charge). 

We first find a linear response function ${\cal C}(t)$ that relates the current through the junction to the phase bias $\delta \varphi(t)$. In the considered gauge, the current operator at the position of the junction is given by $J(\varphi) = 2e \partial_\varphi H(\varphi)$. The time-dependent perturbation due to the bias is obtained by expanding $H(\varphi + \delta\varphi(t))$ to the first order in $\delta\varphi(t)$; it can be represented as $H_V \approx J(\varphi)\delta \varphi(t) / 2e$. To find the linear response relation, we consider the difference $\delta J(t) = \langle J(\varphi + \delta\varphi(t))\rangle_t - \langle J(\varphi)\rangle_0$, where $\langle \dots\rangle_{0/t} = \mathrm{Tr}[\dots \rho_{0/t}]$, and $\rho_{0}$ ($\rho_t$) is the density matrix of the system before (after) application of the perturbation. By solving the equations of motion for $\rho_t$ to the first order in $\delta\varphi(t)$, for $\delta J(\omega) = \int dt \, e^{i\omega t} \delta J(t)$ we obtain
\begin{equation}
    \delta J(\omega)=\frac{1}{2e}{\cal{C}}(\omega)\delta\varphi(\omega),
\end{equation}
where the factor of $1/(2e)$ was introduced for convenience, $\delta \varphi(\omega) = \int dt \, e^{i\omega t} \delta\varphi(t)$, and
\begin{equation}\label{eq:app_resp_func}
    {\cal C}(\omega)=2e \left\langle  \partial_\varphi J(\varphi)\right\rangle_0 +{\cal C}^{R}_{JJ}(\omega), 
\end{equation}
The first term in the expression for ${\cal C}(\omega)$ describes the diamagnetic contribution to the response function; it originates from the expansion of the current operator $J(\varphi + \delta \varphi(t))$ to the first order in $\delta \varphi(t)$. The second term, ${\cal{C}}_{JJ}^{R}(\omega)$, results from a first order contribution to $\rho_t$ and is given by a Kubo formula,
\begin{equation}\label{eq:app_kubo_resp}
    {\cal C}^{R}_{JJ}(\omega) = -i\int_0^{+\infty} dt \, e^{i\omega t} \langle[J(\varphi, t), J(\varphi, 0)] \rangle_0 .
\end{equation}
The expression for ${\cal C}(\omega)$ can be further simplified assuming that $\rho_0 \equiv \rho_0(H(\varphi))$ with $\rho_0(E)$ an analytic function of $E$ [\textit{e.g.,} this is the case in thermal equilibrium, but we also allow for non-equilibrium situations]. In this case, the diamagnetic term can be represented as
\begin{equation}\label{eq:app_kubo_deriv}
\left\langle \partial_\varphi J(\varphi)\right\rangle_0 = \partial_\varphi \left\langle J(\varphi)\right\rangle_0 - \mathrm{Tr}\bigl[J(\varphi) \partial_\varphi\rho_0(H(\varphi))\bigr],
\end{equation}
where $\langle J (\varphi)\rangle_0$ is the stationary Josephson current at phase-difference $\varphi$. It is convenient to express the trace on the right hand side as a sum over the many-body eigenstates $|a\rangle$, $|b\rangle$ of $H(\varphi)$ (with energies $E_a$, $E_b$, respectively):
\begin{align}\label{eq:app_tr_simp}
    \mathrm{Tr}\bigl[J(\varphi) \partial_\varphi \rho_0(H(\varphi))\bigr] &= \sum_{a,b} \langle a | J(\varphi) | b \rangle \langle b |\partial_\varphi \rho_0(H(\varphi))|a \rangle \notag \\
    &= \frac{1}{2e}\sum_{a,b}|\langle a|J(\varphi) |b\rangle|^2 \frac{\rho_0(E_b) - \rho_0(E_a)}{E_b - E_a}.
\end{align}
Here, the second equality can be verified straightforwardly by expanding $\rho_0(H(\varphi))$ in powers of $H(\varphi)$ and then computing the derivative with respect to $\varphi$. The final expression in Eq.~\eqref{eq:app_tr_simp} coincides with ${\cal C}_{JJ}^R(0) / (2e)$, as can be easily checked using Eq.~\eqref{eq:app_kubo_resp} \footnote{If there is a bound state present at the junction, then it is also necessary to assume that the frequency $\omega$ exceeds the parity lifetime of the bound state.}. Then, by combining Eqs.~\eqref{eq:app_resp_func} and \eqref{eq:app_kubo_deriv}, we obtain
\begin{equation}\label{eq:app_resp_func_gen}
    {\cal C}(\omega) = 2e \partial_\varphi \langle J(\varphi) \rangle_0 + \bigl[{\cal C}^R_{JJ}(\omega) - {\cal C}^R_{JJ}(0)\bigr].
\end{equation}
It can be easily shown that this equation holds more generally whenever the single-particle distribution function depends on energy alone.

As a next step, we relate the response function ${\cal C}(\omega)$ to the admittance $Y(\omega)$. According to the Josephson relation, $\delta \varphi(\omega) = 2e i V(\omega) / \omega$. Consequently, $Y(\omega) \equiv \delta J(\omega) / V(\omega) = i {\cal C}(\omega) / \omega$ and its dissipative and non-dissipative parts are  given by
\begin{align}\label{eq:app_reY_g}
    \mathrm{Re}\,Y(\omega) &= -\frac{1}{\omega}\mathrm{Im}\,{\cal C}_{JJ}^R(\omega),\\\label{eq:app_imY_g}
    \mathrm{Im}\,Y(\omega) &= \frac{2e}{\omega} \partial_\varphi \langle J(\varphi)\rangle_0 + \frac{1}{\omega} \mathrm{Re}\bigl[{\cal C}^R_{JJ}(\omega) - {\cal C}^R_{JJ}(0)\bigr],
\end{align}
respectively. The linear response function ${\cal C}^R_{JJ}(\omega)$ is analytic in the upper-half complex plane of $\omega$ and thus satisfies the Kramers-Kronig relation,
\begin{equation}\label{eq:app_KK_r}
    \mathrm{Re}\,{\cal C}^R_{JJ}(\omega) = \frac{1}{\pi} \fint_{-\infty}^{+\infty} \frac{\mathrm{Im}\,{\cal C}^R_{JJ}(\omega^\prime)}{\omega^\prime - \omega}d\omega^\prime.
\end{equation}
We conclude from Eqs.~\eqref{eq:app_reY_g}--\eqref{eq:app_KK_r} that
\begin{equation}\label{eq:app_im_adm_gen}
    \mathrm{Im}\,Y(\omega) = \frac{2e}{\omega}\partial_\varphi\langle J(\varphi) \rangle_0 - \frac{2\omega}{\pi} \fint_{0}^{+\infty} \frac{\mathrm{Re}\,Y(\omega^\prime)}{\omega^{\prime 2} - \omega^2}d\omega^\prime,
\end{equation}
where we used $\mathrm{Re}\,Y(\omega^\prime) =\mathrm{Re}\,Y(-\omega^\prime)$ to simplify the final expression.

At $T = 0$ the stationary Josephson current is related to the ground state energy of the junction $E_\mathrm{gs}(\varphi, B)$ through $\langle J(\varphi) \rangle_0 = 2e \partial_\varphi E_\mathrm{gs}(\varphi, B)$. Using this relation in Eq.~\eqref{eq:app_im_adm_gen} and taking the difference between $\mathrm{Im}\,Y(\omega)$ at the magnetic fields $B$ and $B_\mathrm{c}$, we arrive at Eqs.~\eqref{eq:im_adm}--\eqref{eq:im_adm_v} of the main text.

\subsection{Scattering matrix of the Majorana modes \label{sec:app_scat_mat}}

Equation \eqref{eq:deltaEgs} indicates that the contribution to the ground state energy from the states of the continuous spectrum can be extracted from the quasiparticle scattering matrix $S(E)$. In this section, we approximately find $S(E)$ at $E \sim E_\mathrm{gap}$ using the low-energy theory of Secs.~\ref{sec:model}--\ref{sec:away}.  To do that, we examine the eigenstates of the single-particle Hamiltonian $\hat{H}$ [see Eq.~\eqref{eq:app_singlep_h}].
Each energy eigenvalue $E > E_\mathrm{gap}$ is two-fold degenerate. The two corresponding (improper) eigenstates can be chosen as scattering states:
\begin{align}
    \psi_R^\mathrm{sc}(x|E) &=
    \begin{cases}
    \psi_R(x|E) + S_{-+}(E) \psi_L(x|E),    & x < 0,\\
    S_{++}(E)\psi_R(x|E),                   & x > 0,
    \end{cases} \\
    \psi_L^\mathrm{sc}(x|E) &=
    \begin{cases}
    S_{--}(E)\psi_L(x|E),                   & x < 0,\\
    \psi_L(x|E) + S_{+-}(E) \psi_R(x|E),    & x > 0.
    \end{cases}
\end{align}
Here, $\psi_{R/L}(x|E)$ describes a right-/left-propagating wave [see Eq.~\eqref{eq:rl-wfs}], and $S_{\pm\pm}(E)$ are the entries of the scattering matrix. The parameters $S_{\pm\pm}(E)$ are found by matching the eigenfunctions across $x = 0$ using $\psi_{R/L}^\mathrm{sc}(0^+|E) = {\cal T}\psi_{R/L}^\mathrm{sc}(0^-|E)$, where the transfer matrix ${\cal T}$ is defined in Eq.~\eqref{eq:app_match_0}. We find
\begin{equation}\label{eq:app_S_matrix}
    S(E) =
    \begin{pmatrix}
    S_{++}(E) & S_{+-}(E)\\
    S_{-+}(E) & S_{--}(E)
    \end{pmatrix} = 
    \frac{1}{(\mathrm{u}_E^2 - \mathrm{v}_E^2) +  2i\mathrm{u}_E\mathrm{v}_E \sgn (M\cdot g) \sqrt{1 - \tau}}
    \begin{pmatrix}
    (\mathrm{u}_E^2 - \mathrm{v}_E^2)\sqrt{\tau} & -\sqrt{1 - \tau}\sgn{g}\\
    \sqrt{1 - \tau}\sgn{g} & (\mathrm{u}_E^2 - \mathrm{v}_E^2)\sqrt{\tau}
    \end{pmatrix},
\end{equation}
where the Bogoliubov amplitudes $\mathrm{u}_\mathrm{E},\mathrm{v}_\mathrm{E}$ are defined in Eq.~\eqref{eq:app_bogoliubov}. The scattering matrix at negative energies $E < -E_\mathrm{gap}$ can be obtained from Eq.~\eqref{eq:app_S_matrix} (which is valid at $E > E_\mathrm{gap}$) using the particle-hole symmetry; the latter implies that $S(-E) = S^\star(E)$. Taking the determinant of $S(E)$ given by Eq.~\eqref{eq:app_S_matrix}, we obtain Eq.~\eqref{eq:detS} of the main text. 

\subsection{Estimate of $\delta\mathrm{Im}\,Y_v(\omega)$ \label{sec:app_imYv}}

In this section, we estimate $\delta\mathrm{Im}\,Y_v(\omega)$ [see Eq.~\eqref{eq:im_adm_v}] under the assumptions that $T = 0$ and $E_\mathrm{gap} \ll \omega \ll \Delta$, where $\Delta$ is the proximity-induced pairing potential in the quantum wire. For simplicity, we also assume that $1 - \tau$ is a number of the order of unity in the estimates below. 

The contribution $\delta\mathrm{Im}\,Y_v(\omega)$ can be found by computing the integral over $\omega^\prime$ in Eq.~\eqref{eq:im_adm_v} with the help of scaling relations \eqref{eq:scY1} and \eqref{eq:scY2}.
Thus, we must compute the integral
\begin{equation}
\delta\mathrm{Im}\,Y_v(\omega) = -\frac{2\omega}{\pi} \fint_{0}^{+\infty}\frac{\delta\mathrm{Re}\,Y(\omega^{\prime})}{\omega^{\prime 2}-\omega^{2}}d\omega^{\prime} ,
\end{equation}
with
\begin{equation}\label{eq:app_exact_dReY}
\delta\mathrm{Re}\,Y(\omega^\prime) = C \omega^{\prime \gamma} \Bigl[ f^{(1)}\Bigl( \frac{\omega^\prime}{E_{\mathrm{gap}}} \Bigr) + \Theta(-M \cdot g) \frac{E_{\mathrm{gap}}}{\omega^\prime} f^{(2)}\Bigl( \frac{\omega^\prime}{E_{\mathrm{gap}}} \Bigr) - 1 \Bigr] .
\end{equation}
As will be verified shortly, the integral converges at $\omega^\prime \sim \omega \gg E_\mathrm{gap}$ which allows us to use an asymptotic expression for $\delta\mathrm{Re}\,Y(\omega^\prime)$:
\begin{equation}\label{eq:app_asym}
    \delta \mathrm{Re}\,Y(\omega^\prime) \approx -\pi \sqrt{1 - \tau} C \omega^{\prime\gamma} \sgn (M\cdot g) \frac{E_\mathrm{gap}}{\omega^\prime}|z_\infty|^2
\end{equation}
[this expression follows directly from Eqs.~\eqref{eq:scaling}, \eqref{eq:scaling2} for the scaling functions $f^{(1,2)}$ at $\omega^\prime \gg E_\mathrm{gap}$]. Here, $\gamma$ is the dynamic critical exponent [see Eq.~\eqref{eq:dynexp}] and $|z_\infty|$ is a limit of the transition matrix element $|z(w,\varepsilon)|$ at $w \rightarrow \infty$ [the limit does not depend on $\varepsilon$, see Eqs.~\eqref{eq:matrixnosyms}--\eqref{eq:matrixsymnoI}]. When ${\cal M}_x$ and $\cal{R}$ are both absent, $\gamma = 0$ and $|z_\infty| = 1$. Thus for $\delta \mathrm{Im}\,Y_v$ we obtain
\begin{equation}
    \delta \mathrm{Im}\,Y_v \approx 2C\sqrt{1-\tau}\sgn (M \cdot g) E_\mathrm{gap} \fint_{E_\mathrm{gap}}^{+\infty} \frac{d\omega^\prime}{\omega^\prime}\frac{\omega}{ \omega^{\prime 2} - \omega^2}.
\end{equation}
At $E_\mathrm{gap} \ll \omega^\prime \ll \omega$ the integrand is $\propto 1 / \omega^\prime$ and therefore the integral is logarithmic. Computing it we obtain Eq.~\eqref{eq:imYv} of the main text. Note that the integral converges at $\omega^\prime \sim \omega \gg E_\mathrm{gap}$, justifying the applicability of the asymptotic expression \eqref{eq:app_asym}.
Indeed, the asymptotic expression deviates from the exact expression~\eqref{eq:app_exact_dReY} only near $\omega^\prime \sim E_{\mathrm{gap}}$; the difference may affect the result by an amount of order at most $E_{\mathrm{gap}}/\omega$, and such a correction to Eq.~\eqref{eq:imYv} can be neglected in the leading logarithmic approximation [\textit{i.e.}, when $\ln (\omega / E_\mathrm{gap}) \gg 1$]. We also note that, while Eq.~\eqref{eq:imYv} was derived under the assumption that $1 - \tau$ is of the order of unity, it remains applicable in the opposite limit, $1 - \tau \ll 1$, provided that $\ln (\omega / E_\mathrm{gap}) \gg 1 / \sqrt{1 - \tau}$.

If at least one of the two symmetries is present, then $\gamma = 2$ and $|z_\infty|^2 = 2\lambda + 1$ [with $\lambda = 1$ if ${\cal M}_x$ is a symmetry of the system, cf.~Eqs.~\eqref{eq:matrixsymI} and \eqref{eq:matrixsymnoI}]. 
In this case, we find
\begin{equation}
    \delta \mathrm{Im}\,Y_v \approx 2C\sqrt{1-\tau}(2\lambda + 1)  \sgn (M  \cdot g) E_\mathrm{gap} \fint_{E_\mathrm{gap}}^{+\infty} \omega^\prime d\omega^\prime \frac{\omega}{\omega^{\prime 2} - \omega^2}.
\end{equation}
The integrand behaves as $\propto 1 / \omega^\prime$ for $\omega^\prime \gg \omega \gg E_\mathrm{gap}$ [notice the difference in the relevant domain of $\omega^\prime$ compared to the case of $\gamma = 0$]. Thus the integral is logarithmically divergent at the upper limit. The ultraviolet cutoff is provided by the proximity induced pairing potential $\Delta$. We obtain
\begin{align}
    \delta \mathrm{Im}\,Y_v \approx  C\sqrt{1-\tau}(2\lambda + 1) \sgn (M  \cdot g) E_\mathrm{gap}\,\omega \ln \Bigl[\frac{\Delta}{\omega}\Bigr].
\end{align}
Notice that this leading-order expression is an analytic function of $B - B_\mathrm{c}$: $\delta \mathrm{Im}\,Y_v \propto B - B_\mathrm{c}$. We conclude that, in the presence of symmetries, the contribution $\delta \mathrm{Im}\,Y_v$ is negligible in comparison with a logarithmically-larger nonanalytic contribution $\delta \mathrm{Im}\,Y_i \propto (B - B_\mathrm{c}) \ln \bigl[ B_\mathrm{c} / (B - B_\mathrm{c}\bigr]$, as discussed in Sec.~\ref{sec:non_diss}.

\subsection{Estimate for  splitting $\mathrm{Im}\,\widetilde{Y}(\omega) - \mathrm{Im}\,Y(\omega)$\label{sec:app_splitting}}

The goal of this Appendix is to elucidate how the difference
\begin{equation}
    \bar{\delta}\mathrm{Im}\,Y(\omega) = \mathrm{Im}\,\widetilde{Y}(\omega) - \mathrm{Im}\,Y(\omega) 
\end{equation}
depends on the magnetic field $B$ in the vicinity of the critical field $B_\mathrm{c}$; recall that $Y(\omega)$ and $\widetilde{Y}(\omega)$ denote the admittances of the junction in states with 0 and 1 quasiparticles trapped at the bound state, respectively [we use the notation $\bar{\delta}$ to distinguish $\bar{\delta}\mathrm{Im}\,Y(\omega)$ from the quantity $\delta\mathrm{Im}\,Y(\omega)$ introduced in Eq.~\eqref{eq:im_adm}]. Below we assume that $M\cdot g < 0$, which guarantees the existence of the bound state at the junction [see Sec.~\ref{sec:away}]. For simplicity, we concentrate on the limit $T = 0$ and thus assume that above-the-gap ($E > E_\mathrm{gap}$) excitations are absent. 

Using Eq.~\eqref{eq:app_im_adm_gen} it is possible to represent $\bar{\delta}\mathrm{Im}\,Y(\omega)$ as
\begin{equation}\label{eq:app_pois_div}
    \bar{\delta}\mathrm{Im}\,Y(\omega) = \bar{\delta}\mathrm{Im}\,Y_i(\omega) + \bar{\delta}\mathrm{Im}\,Y_v(\omega),
\end{equation}
where $\bar{\delta}\mathrm{Im}\,Y_i(\omega)$ ($\bar{\delta}\mathrm{Im}\,Y_v(\omega)$) corresponds to the difference in the first (second) term of Eq.~\eqref{eq:app_im_adm_gen} between the two considered states. We first discuss the magnetic field dependence of the contribution $\bar{\delta}\mathrm{Im}\,Y_i(\omega)$. This contribution can be expressed as 
\begin{equation}
    \bar{\delta}\mathrm{Im}\,Y_i(\omega) = \frac{2e}{\omega} \partial_\varphi\bigl\{ \bar{\delta}\langle J(\varphi)\rangle\bigl\}.
\end{equation}
Here $\bar{\delta}\langle J(\varphi)\rangle$ denotes the change in the Josephson current upon occupying the bound state.
$\bar{\delta}\langle J(\varphi)\rangle$ is related to the bound state energy $E_\tau$ via $\bar{\delta}\langle J(\varphi)\rangle = 2e \partial_\varphi E_\tau$. Then, using Eq.~\eqref{eq:bound} we find
\begin{equation}\label{eq:app_pois_Yi}
    \bar{\delta}\mathrm{Im}\,Y_i(\omega) = \frac{(2e)^2}{\omega}E_\mathrm{gap}\partial^2_\varphi \sqrt{\tau}.
\end{equation}
Consequently, close to the critical point
\begin{equation}
    \bar{\delta}\mathrm{Im}\,Y_i \propto B - B_\mathrm{c}.
\end{equation}

Next, with the help of Eqs.~\eqref{eq:app_contributions}, \eqref{eq:app_scaling_2}, and \eqref{eq:app_scaling_3} the second term in Eq.~\eqref{eq:app_pois_div} can be represented as
\begin{align}\label{eq:app_imYv2_int_2}
    \bar{\delta} \mathrm{Im}\,Y_v(\omega) = -\frac{2C\omega }{\pi} E_\mathrm{gap} \fint_0^{+\infty} \frac{\omega^{\prime \gamma - 1} d \omega^\prime}{\omega^{\prime 2} - \omega^{2}} \Bigl[f^{(3)}\Bigl(\frac{\omega^\prime}{E_\mathrm{gap}}\Bigr) - f^{(2)}\Bigl(\frac{\omega^\prime}{E_\mathrm{gap}}\Bigr)\Bigr],
\end{align}
where in the expression for $f^{(2)}$ we take $\tilde{n} = 0$ and $p_\tau = 0$ [see Eq.~\eqref{eq:app_scaling_2}], while in the expression for $f^{(3)}$ we take $\tilde{n} = 0$ and $p_\tau = 1$ [see Eq.~\eqref{eq:app_scaling_3}]. The character of the magnetic field dependence of $\bar{\delta} \mathrm{Im}\,Y_v$ depends on the symmetry of the system. By estimating the integral over $\omega^\prime$ under the assumption $\omega \ll E_\mathrm{gap}$ we obtain
\begin{equation}\label{eq:app_im_adm_v_pois}
    \bar{\delta} \mathrm{Im}\,Y_v \propto E_\mathrm{gap}^{\gamma + 1}.
\end{equation}
If ${\cal M}_x$ and $\cal{R}$ are both absent, then $\gamma = 0$ and $\bar{\delta} \mathrm{Im}\,Y_v \propto B - B_\mathrm{c}$. In this case, the dependence of $\bar{\delta}\mathrm{Im}\,Y_v$ on the magnetic field is similar to that of $\bar{\delta}\mathrm{Im}\,Y_i$. If at least one of the two symmetries is present, then $\gamma = 2$ and $\bar{\delta} \mathrm{Im}\,Y_v \propto (B - B_\mathrm{c})^3$ is subleading in comparison with $\bar{\delta}\mathrm{Im}\,Y_i$.

Summarizing Eqs.~\eqref{eq:app_pois_Yi} and \eqref{eq:app_im_adm_v_pois} we conclude that 
\begin{equation}
    \mathrm{Im}\,\widetilde{Y}(\omega) - \mathrm{Im}\,Y(\omega) \propto B - B_\mathrm{c}
\end{equation}
regardless of the symmetry of the system. This justifies the proportionality $\omega_\mathrm{sp} \propto \Theta(-M\cdot g) (B - B_\mathrm{c})$ presented in Sec.~\ref{sec:signatures}.

\section{Symmetry ${\cal R}$ in a more realistic model of a topological junction\label{sec:app_R}}

The goal of this Appendix is to highlight that the antiunitary symmetry ${\cal R}$ is not tied to the particular simple one-dimensional model considered in Sec.~\ref{sec:micro}, but may also be present in more sophisticated and realistic models which account for the wire's three-dimensional geometry and orbital effects of the external magnetic field. 

\begin{figure}[t]
  \begin{center}
    \includegraphics[scale = 1.0]{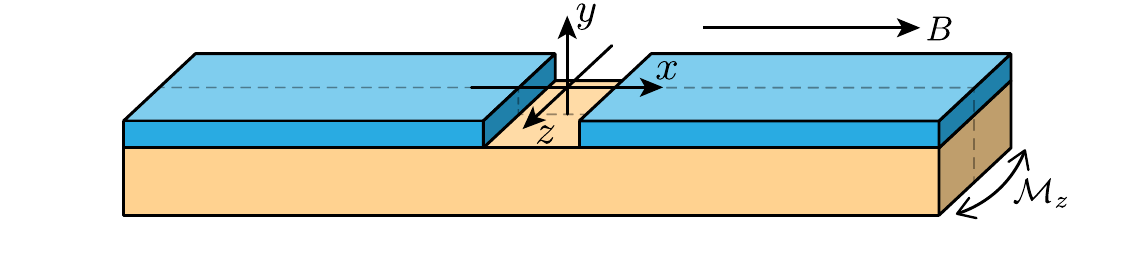}
    \caption{Sketch of the setup considered in Appendix~\ref{sec:app_R}. A semiconducting nanowire (yellow) is covered in two segments by superconducting shells (blue) so that a Josephson junction is formed. The junction is placed in an external uniform magnetic field $B$ directed along the $x$-axis. If the nanowire and the superconducting shells are symmetric under mirror reflection in the $z$-direction, the system has an antiunitary symmetry ${\cal R} = {\cal M}_z {\cal T}$ [see discussion around Eq.~\eqref{eq:app_assumptions}].}
    \label{fig:3D_wire}
  \end{center}
\end{figure}

We consider a Josephson junction formed by two proximitized segments of a semiconducting nanowire 
[see Fig.~\ref{fig:3D_wire}]. The junction is placed in an external uniform magnetic field $\bm{B}$ aligned with the wire's axis. The mean-field many-body Hamiltonian describing the electrons in the system is given by 
\begin{equation}
    H = \frac{1}{2}\int d^3\bm{r}\,\Psi^\dagger (\bm{r}) \hat{H} \Psi(\bm{r}),
\end{equation}
where $\Psi = \bigl(\psi_\uparrow,\,\psi_\downarrow,\,\psi^\dagger_\downarrow,\,-\psi^\dagger_\uparrow \bigr)^T$ and $\psi_\sigma$ is an annihilation operator of electrons with spin $\sigma$. The Bogoliubov-de Gennes Hamiltonian $\hat{H}$ is given by (see, \textit{e.g.}, Ref.~\cite{nijholt2016})
\begin{equation}\label{eq:app_3D_wire}
    \hat{H} = \Bigl[\frac{\bm{p}^2}{2m} - \mu + U(\bm{r}) + v (\sigma_z p_x - \sigma_x p_z)  \Bigr] \tau_z + \Delta (\bm{r}) \tau_x - \frac{1}{2}g \mu_B B \sigma_x,
\end{equation}
with Pauli matrices $\sigma_{x,y,z}$ ($\tau_{x,y,z}$) acting in the spin (Nambu) space.
Here $U(\bm{r})$ is the potential energy of an electron in the wire; $U(\bm{r})$ incorporates the wire's confining potential, the potential of the electric field arising from the lack of inversion symmetry in the $y$-direction, and the scattering potential at the junction; $\bm{p} = -i\nabla + e \bm{A} \tau_z /c$ is the canonical momentum, and $\bm{A}$ is the vector potential. The term $g \mu_B B \sigma_x / 2$ describes the Zeeman effect. Finally, $\Delta(\bm{r})$ is the proximity-induced pairing potential; we assume that the gauge is fixed in such a way that $\Delta(\bm{r})$ is real.

Let us consider the wire and superconducting shells symmetric under mirror reflection $z\rightarrow -z$, \textit{i.e.,}
\begin{equation}\label{eq:app_assumptions}
    U(x,y,z) = U(x,y,-z),\quad\quad \Delta(x,y,z) = \Delta(x,y,-z).
\end{equation}
Let us also assume that there is no phase bias applied to the superconducting leads. Now, suppose for a moment that the external magnetic field is switched off. In this case, Hamiltonian \eqref{eq:app_3D_wire} is symmetric under the mirror reflection ${\cal M}_z = \exp{(i\pi \sigma_z / 2)} {\cal P}_z$ and under time-reversal ${\cal T} = i\sigma_y {\cal K}$ (where ${\cal P}_z z = -z$ and ${\cal K}$ is complex conjugation). The external magnetic field $B \neq 0$ breaks both ${\cal M}_z$ and ${\cal T}$ individually, but preserves their combination ${\cal R} = {\cal M}_z {\cal T}$. To see this, note that the external magnetic field changes sign under time reversal. It changes sign again under ${\cal M}_z$, as follows from the fact that $\bm{B}$ is a pseudovector parallel to the reflection plane. Thus the Hamiltonian is invariant under the action of ${\cal R}$. This demonstrates that the symmetry ${\cal R}$ may indeed be present in realistic devices without fine-tuning of the system's parameters.

Notice that the representation ${\cal R} = \exp{(i\pi \sigma_z / 2)} {\cal P}_z {\cal T}$ differs from the one in Eq.~\eqref{eq:symmsR} by the presence of the parity operator ${\cal P}_z$. Including ${\cal P}_z$ in the definition of ${\cal R}$ is essential when the three-dimensional geometry of the wire is taken into account. Indeed consider, for example, the spin-orbit coupling term $v \sigma_x (-i\partial_z)$. It is invariant under ${\cal T}$ and transforms into minus itself under the spin-rotation $\exp{(i\pi \sigma_z / 2)}$. The parity operator ${\cal P}_z$ changes the sign of this term again, ensuring that the term is invariant under ${\cal R}$.

The model discussed in this Appendix allows us to identify an important class of perturbations that break symmetry ${\cal R} = {\cal M}_z {\cal T}$, which are not captured by the simple model of Sec.~\ref{sec:micro}. These are violations of the mirror symmetry ${\cal M}_z$ by the geometry of the device. As an example, ${\cal R}$ is broken if the superconducting shells cover the wire asymmetrically with respect to $z \rightarrow -z$. This is often the case in realistic devices, \textit{e.g.}, in hexagonal InAs nanowires with superconducting shells covering only two of the six facets \cite{krogstrup2015}. 

Finally, we note that the antiunitary symmetry ${\cal R}$ is closely related to the chiral symmetry of topological quantum wires discussed, \textit{e.g.}, in Refs.~\cite{nijholt2016, winkler2019}. Let us combine ${\cal R} = {\cal M}_z {\cal T}$ with the particle-hole symmetry $-i\tau_y {\cal T}$ of the Bogoliubov-de Gennes Hamiltonian \eqref{eq:app_3D_wire}. This defines a unitary operation ${\cal C} = \tau_y \sigma_z {\cal P}_z$. If Hamiltonian \eqref{eq:app_3D_wire} is symmetric under ${\cal R}$, it also satisfies
\begin{equation}
    {\cal C} \hat{H} {\cal C}^\dagger = - \hat{H}.
\end{equation}
Thus ${\cal C}$ is a chiral symmetry of the model. It coincides with symmetry ${\cal C}^\prime$ of Ref.~\cite{nijholt2016} and with symmetry ${\cal C}$ of Ref.~\cite{winkler2019}.

\section{Evaluation of the parameters of the low-energy theory from the microscopic Hamiltonian}

In this Appendix, we describe how the parameters of the critical theory can be computed by performing a projection of the microscopic Hamiltonian onto the low-energy subspace. We start by presenting a general framework for how the projection is done. We then apply this framework to analyze different cases [see Secs.~\ref{sec:app_1} and \ref{sec:app_zetaz}].

Following Sec.~\ref{sec:micro}, we consider the nanowire Josephson junction described by the Hamilonian $H + V$, where
\begin{equation}\label{eq:app_Hmb}
H = \frac{1}{2}\int dx
\begin{pmatrix}
\chi^T(x) & \eta^T(x)
\end{pmatrix}
\begin{pmatrix}
\hat{H}_\chi (-i\partial_x)   &   0     \\
0         &   \hat{H}_\eta (-i\partial_x)
\end{pmatrix}
\begin{pmatrix}
\chi(x)    \\
\eta(x)
\end{pmatrix}, \quad \hat{H}_{\chi/\eta}(p) = v p \zeta_z + (B \mp \Delta)\zeta_y,
\end{equation}
[see Eq.~\eqref{eq:HM}; we remind that $\chi$ are the low-energy modes and $\eta$ are the high-energy modes] and $V$ describes perturbations to $H$ that originate, \textit{e.g.}, due to a finite phase bias, scattering at the junction, etc. All of the perturbations $V$ that we focus on in this Appendix can be represented as
\begin{equation}\label{eq:perturbation}
V = \frac{1}{2}\int dx
\begin{pmatrix}
\chi^T(x) & \eta^T(x)
\end{pmatrix}
\begin{pmatrix}
0    &   i \hat{V}_{\chi\eta}(x) \\
-i \hat{V}^\dagger_{\chi\eta}(x)   & 0
\end{pmatrix}
\begin{pmatrix}
\chi(x)    \\
\eta(x)
\end{pmatrix},
\end{equation}
where the functional form and the matrix structure of $\hat{V}_{\chi\eta}(x)$ depend on the origin of the perturbation.

To project the Hamiltonian $H + V$ onto the low-energy subspace, we perform a unitary Schrieffer-Wolff transformation that removes the coupling between the low-energy ($\chi)$ and the high-energy ($\eta$) Majorana modes:
\begin{equation}\label{eq:app_sw}
\begin{pmatrix}
\chi    \\
\eta
\end{pmatrix} \rightarrow
\exp
\begin{pmatrix}
0   &   i \hat{W} \\
i \hat{W}^\dagger &   0
\end{pmatrix} \begin{pmatrix}
\chi    \\
\eta
\end{pmatrix}
\end{equation}
(with the operator $\hat{W}$ satisfying $\hat{W} = - \hat{W}^\star$ to preserve the anti-commutation relations of the Majorana fields). The operator $\hat{W}$ can be constructed perturbatively in $V$. By requiring that the coupling between $\chi$ and $\eta$ is removed to the first order in $V$, we obtain the lowest-order equation for $\hat{W}$, which we express in the momentum domain:
\begin{equation}\label{eq:app_S}
    \hat{H}_\chi(k) \hat{W}_{kp} - \hat{W}_{kp} \hat{H}_\eta(p) = -\frac{1}{L} \hat{V}^{k-p}_{\chi\eta}.
\end{equation}
Here, $\hat{W}_{kp} = \langle k|\hat{W}|p\rangle$ [we define the plane-wave states as  $\langle x|q\rangle = e^{iqx}/\sqrt{L}$, where $L$ is the system size] and $\hat{V}_{\chi\eta}^{q} = \int dx \, e^{-iqx} \hat{V}_{\chi\eta}(x)$.  An effective low-energy Hamiltonian $\hat{H}_\mathrm{eff}$ can then be found by taking a $\chi\chi$-component of the transformed (single-particle) Hamiltonian. To the second order in $V$, it is given by
\begin{equation}\label{eq:app_Heff_S}
    \hat{H}_\mathrm{eff}^{kp} \approx \hat{H}_\chi(k)\delta_{kp} - \frac{1}{2} \bigl(\hat{W}\hat{V}^\dagger_{\chi\eta} + \hat{V}_{\chi\eta} \hat{W}^\dagger\bigr)_{kp},   
\end{equation}
where $\delta_{kp}$ is a Kronecker delta.

The low-energy current operator $I$ can be obtained by applying the Schrieffer-Wolff transformation [Eq.~\eqref{eq:app_sw}] to the microscopic current operator $J(x = 0)$ [see Eq.~\eqref{eq:current_micro}] and then taking its $\chi\chi$-component. In terms of $\hat{W}$, we find (in the momentum domain)
\begin{equation}\label{eq:app_current_S}
    \hat{I}_{kp} \approx - \frac{e}{L} \frac{k + p}{2m}\mathbbm{1} + \frac{ev}{L} \sum_q\Bigl\{\hat{W}_{kq}\zeta_z  + \zeta_z  \bigl(\hat{W}^\dagger\bigr)_{qp}\Bigr\},
\end{equation}
where $\hat{I}$ is the single-particle representation of the low-energy current operator $I$ [we remind that $\hat{I}(x)$ is defined in such a way that $I = \frac{1}{2} \int dx \chi^T(x) \hat{I}(x)\chi(x)$].

In the following sections we apply Eqs.~\eqref{eq:app_S}--\eqref{eq:app_current_S} to compute the parameters of the low-energy theory in the presence of different perturbations $V$ [we assume that only one type of perturbation is present at a time]. We focus on two classes of perturbations: $\hat{V}_{\chi\eta}(x) = V_0(x) \mathbbm{1}$ [see Sec.~\ref{sec:app_1}] and $\hat{V}_{\chi\eta}(x) = V_z(x) \zeta_z$ [see Sec.~\ref{sec:app_zetaz}], where $V_0(x)$ and $V_z(x)$ are real functions. These two types cover all of the examples considered in Sec.~\ref{sec:micro}. Below we will always assume that the magnetic field is tuned to the vicinity of the topological transition, $|B - B_\mathrm{c}| \ll \Delta$, where $B_\mathrm{c} = \Delta$ in our model.

\subsection{Perturbations of the form  $\hat{V}_{\chi\eta}(x) = V_0(x) \mathbbm{1}$ \label{sec:app_1}}

First, we consider perturbations of the form $\hat{V}_{\chi\eta}(x) = V_0(x) \mathbbm{1}$. This type of perturbation can describe a scattering potential at the junction, $u(x)$, or a chemical potential $\mu \neq 0$ [see Secs.~\ref{sec:app_scat} and \ref{sec:app_mu}, respectively]. By solving Eq.~\eqref{eq:app_S}, we find the corresponding transformation matrix $\hat{W}$ to the first order in $V_0$:
\begin{equation}
    \hat{W}_{kp} \approx  \frac{1}{L} \frac{v(k+p)\zeta_z + 2\Delta \zeta_y}{4\Delta^2 + v^2(p^2 - k^2)} V_0^{k-p}.
\end{equation}
Then, according to Eqs.~\eqref{eq:app_Heff_S}, \eqref{eq:app_current_S}, the low-energy Hamiltonian and current operator are given by
\begin{align}\label{eq:app_V0_H}
    \hat{H}^{kp}_\mathrm{eff} &\approx \hat{H}_{\chi}(k)\delta_{kp} - \frac{1}{2L^2}\sum_q \Bigl(\frac{v(k+q)\zeta_z + 2\Delta \zeta_y}{4\Delta^2 + v^2(q^2 - k^2)} + \frac{v(p+q)\zeta_z + 2\Delta \zeta_y}{4\Delta^2 + v^2(q^2 - p^2)}\Bigr) V_0^{k-q} V_0^{q-p},\\
    \hat{I}^{kp} &\approx -\frac{e}{L} \frac{k + p}{2m}\mathbbm{1} + \frac{e v}{L^2}\sum_q \Bigl(\frac{v(k+q)\mathbbm{1} + 2i\Delta \zeta_x}{4\Delta^2 + v^2(q^2 - k^2)} V_0^{k-q} + \frac{v(p+q)\mathbbm{1} - 2i\Delta \zeta_x}{4\Delta^2 + v^2(q^2 - p^2)} V_0^{q-p} \Bigr).\label{eq:app_V0_I}
\end{align}
Now, the parameters of the low-energy theory ($v_{R/L}$, $g$, $\alpha$, $\kappa_{ij}$) can be found by performing a gradient expansion in the expressions for $\hat{H}^{kp}_\mathrm{eff}$ and $\hat{I}^{kp}$. We do that for the concrete examples of $V_0(x)$.

\subsubsection{Scattering potential\label{sec:app_scat}}

The scattering potential is described by $V_0(x) \equiv u(x)$, where the function $u(x)$ is localized around $x = 0$ on some typical length scale $l_u$ [we remind that $u(x)$ respects the symmetry $\cal{R}$ but breaks ${\cal M}_x$ if $u(x)\neq u(-x)$]. We will assume that $l_u$ is small compared to the relevant wavelengths, \textit{i.e.}, the momenta $k$ and $p$ satisfy $kl_u, pl_u \ll 1$. We will also assume that $k\xi, p\xi \ll 1$, where $\xi = v / \Delta$ is the coherence length. Then, expanding Eqs.~\eqref{eq:app_V0_H}, \eqref{eq:app_V0_I} to the lowest order in these small parameters  we find
\begin{align}\label{eq:app_g_scat}
    \hat{H}^{kp}_\mathrm{eff} &\approx \hat{H}_{\chi}(k)\delta_{kp} - \frac{\zeta_y}{2L}\int\frac{dq}{2\pi}\frac{4\Delta}{4\Delta^2 + v^2q^2} |u_{q}|^2,\\
    \hat{I}^{kp} &\approx -\frac{e}{L} \frac{k + p}{2m}\mathbbm{1} + \frac{e}{L} v^2(k+p)\mathbbm{1}\int \frac{dq}{2\pi} \frac{u_q - q\partial_q u_q}{4\Delta^2 + v^2q^2} + \frac{e}{L} v(k+p)\zeta_x \int \frac{dq}{2\pi} \frac{2i\Delta \partial_q u_q}{4\Delta^2 + v^2q^2},\label{eq:app_alpha_scat}
\end{align}
where we changed sums over $q$ to integrals and where $u_q = \int dx \, e^{-iqx}u(x)$. The second term in Eq.~\eqref{eq:app_g_scat} is independent of $k$ and $p$. Therefore, it is proportional to $\delta(x)$ in real space. We then find the expression for the parameter $g$ in Eq.~\eqref{eq:g_u} from Eq.~\eqref{eq:app_g_scat}.  Next, $(k + p) / L$ corresponds to $-i\bigl(\delta(x)\overrightarrow{\partial_x} - \overleftarrow{\partial_x} \delta(x)\bigr)$ in real space. Then, from Eq.~\eqref{eq:app_alpha_scat} it follows that $\alpha = 0$, while $\kappa_0, \kappa_x$ are given by Eqs.~\eqref{eq:kappa_0_u}, \eqref{eq:kappa_x_u} of the main text, respectively, and $\kappa_y = 0$.

\subsubsection{Chemical potential\label{sec:app_mu}}
In Sec.~\ref{sec:micro}, we assumed that $\mu = 0$. The effects of $\mu \neq 0$ can be analyzed perturbatively at $\mu \ll \Delta$ by taking $V_0(x) = -\mu$ in Eqs.~\eqref{eq:app_V0_H}, \eqref{eq:app_V0_I}. In this case, $V_0^q = -L\delta_{q,0} \mu$ and therefore
\begin{align}
    \hat{H}^{kp}_\mathrm{eff} &\approx \hat{H}_{\chi}(k)\delta_{kp} - \frac{\mu^2}{2\Delta}\delta_{kp}\zeta_y - \frac{\mu^2}{2\Delta^2} vk \, \zeta_z \delta_{kp},\\
    \hat{I}^{kp} &\approx -e \frac{k + p}{2m}\frac{\mathbbm{1}}{L}  -\frac{e}{L} v(k+p)\frac{v\mu}{2\Delta^2}\mathbbm{1}.
\end{align}
Notice the second term in the expression for $\hat{H}^{kp}_\mathrm{eff}$. It describes a shift of the critical field from $B_\mathrm{c} = \Delta$ to $B_\mathrm{c} \approx \Delta + \mu^2/2\Delta$. This perturbative expression for $B_\mathrm{c}$, valid when $\mu \ll \Delta$, is in agreement with the exact formula $B_\mathrm{c} = \sqrt{\Delta^2 + \mu^2}$. The third term describes the renormalization of the velocity of the Majorana modes due to $\mu \neq 0$.  The parameters of the low-energy theory corresponding to $V_0(x) = -\mu$ are
\begin{equation}
    v_{R/L}= \Bigl( 1 - \frac{\mu^2}{2\Delta^2} \Bigr) v,\quad g=0,\quad \alpha=0,\quad \kappa_{x,y} = 0,\quad \kappa_0 \approx \frac{1}{2m} + \frac{v^2\mu}{2\Delta^2}.
\end{equation}
They are consistent with Table~\ref{tab:class} as $\mu$ does not break $\cal{R}$ or ${\cal M}_x$.

\subsection{Perturbations of the form $\hat{V}_{\chi\eta}(x) = V_z(x) \zeta_z$ \label{sec:app_zetaz}}

Perturbations of the form $\hat{V}_{\chi\eta}(x) = V_z(x) \zeta_z$ can describe a phase bias, the presence of a magnetic barrier at the junction, or the influence of a magnetic field component along the spin-orbit coupling axis [see Secs.~\ref{sec:app_phase}, \ref{sec:app_mbarrier}, and \ref{sec:app_bz} respectively]. For this type of perturbation, the transformation matrix is given by
\begin{equation}
    \hat{W}_{kp} \approx \frac{1}{L} \frac{v(k+p)\mathbbm{1}-2i\Delta \zeta_x}{4\Delta^2 + v^2(p^2 - k^2)} V_z^{k-p}.
\end{equation}
Using this expression in Eqs.~\eqref{eq:app_Heff_S}, \eqref{eq:app_current_S} we find
\begin{align}\label{eq:app_Vz_H}
    \hat{H}^{kp}_\mathrm{eff} &\approx \hat{H}_{\chi}(k)\delta_{kp} - \frac{1}{2L^2}\sum_q \Bigl(\frac{v(k + q) \zeta_z - 2\Delta \zeta_y}{4\Delta^2 +v^2(q^2-k^2)} + \frac{v(p + q) \zeta_z - 2\Delta \zeta_y}{4\Delta^2 +v^2(q^2-p^2)}\Bigr) V_z^{k-q}V_z^{q - p},\\
    \hat{I}^{kp} &\approx  -\frac{e}{L} \frac{k + p}{2m}\mathbbm{1}  + \frac{ev}{L^2}\sum_q \Bigl(\frac{v(k + q)\zeta_z-2\Delta\zeta_y}{4\Delta^2 + v^2(q^2 - k^2)} V_z^{k - q} + \frac{v(p + q) \zeta_z-2\Delta\zeta_y}{4\Delta^2 + v^2(q^2 - p^2)} V_z^{q - p} \Bigr).\label{eq:app_Vz_I}
\end{align}
We now perform a gradient expansion in these formulae to obtain the  parameters of the low-energy theory for different examples of $V_z(x)$.

\subsubsection{Phase difference across the junction\label{sec:app_phase}}

The phase difference across the junction is described by $V_z(x) = -v \varphi \delta(x) / 2$ (it breaks both ${\cal M}_x$ and $\cal{R}$).  In this case, $V_z^q = -v\varphi / 2$ is independent of the momentum $q$. Expanding the summands in Eqs.~\eqref{eq:app_Vz_H}, \eqref{eq:app_Vz_I} to the lowest non-vanishing order in $k\xi, p\xi \ll 1$ and computing the sums over $q$ we find
\begin{align}\label{eq:app_pb}
    \hat{H}^{kp}_\mathrm{eff} \approx \hat{H}_{\chi}(k)\delta_{kp} + \frac{v \varphi^2}{8L} \zeta_y, \qquad
    \hat{I}^{kp} \approx \frac{e v \varphi}{2 L} \zeta_y.
\end{align}
Notice that  we neglected the term $(k+p) \mathbbm{1}/2mL$ in the expression for $\hat{I}^{kp}$. This is allowed when the relevant momenta satisfy $k,p\ll mv\varphi$. We also note that the results in Eq.~\eqref{eq:app_pb} satisfy $\hat{I} = 2e\, \partial \hat{H}_\mathrm{eff} / \partial\varphi$, consistently with the general discussion in Appendix~\ref{sec:app_linresp}. From Eq.~\eqref{eq:app_pb} we obtain $g \approx v \varphi^2 / 8$, $\alpha \approx - v\varphi / 2$, as presented in the main text [see Eq.~\eqref{eq:g_phi} and related discussion]. 

\subsubsection{Magnetic barrier with an antisymmetric magnetization profile\label{sec:app_mbarrier}}

A magnetic barrier with an antisymmetric magnetization profile is described by $V_z(x) \equiv b_z(x)$, where $b_z(x) = -b_z(-x)$ (this perturbation breaks $\cal{R}$ but not ${\cal M}_x$). We will assume that the spatial scale of the barrier, $l_b$, is small compared to the relevant wavelengths, $kl_b, pl_b \ll 1$, and also that $k\xi,p\xi\ll 1$. Then, performing an expansion in Eqs.~\eqref{eq:app_Vz_H}, \eqref{eq:app_Vz_I} to the first order in these small parameters we obtain
\begin{align}
    \hat{H}^{kp}_\mathrm{eff} &\approx \hat{H}_{\chi}(k)\delta_{kp} + \frac{\zeta_y}{2L}\int \frac{dq}{2\pi} \frac{4\Delta }{4\Delta^2 +v^2q^2} |b_z^q|^2, \\ 
    \hat{I}^{kp} &\approx -\frac{e}{L} \frac{k + p}{2m}\mathbbm{1} - \frac{e}{L} i v(p - k)\zeta_y \int \frac{dq}{2\pi} \frac{2i\Delta\partial_q b_z^q}{4\Delta^2 + v^2q^2}.
\end{align}
By converting these expressions to real space [we note that $(p - k)/L$ corresponds to $-i\bigl(\delta(x)\overrightarrow{\partial_x} + \overleftarrow{\partial_x} \delta(x)\bigr)$], we find that $g$ is given by Eq.~\eqref{eq:g_b_z}, $\kappa_0 = 1/2m$, $\kappa_y$ is given by Eq.~\eqref{eq:kappa_y_b_z}, and $\kappa_x = 0$.

\subsubsection{Uniform magnetic field in the direction of the spin-orbit coupling axis\label{sec:app_bz}}

Finally, we use Eqs.~\eqref{eq:app_Vz_H}, \eqref{eq:app_Vz_I} to compute the parameters of the low-energy theory in the case where the magnetic field has a component $B_z \ll \Delta$ along the spin-orbit coupling axis. This perturbation---which breaks both ${\cal M}_x$ and $\cal{R}$---is described by $V_z(x) = B_z$ ($V_z^q = L \delta_{q,0} B_z$ in the momentum domain). We obtain
\begin{align}
    \hat{H}^{kp}_\mathrm{eff} \approx \hat{H}_{\chi}(k)\delta_{kp}  + \frac{B_z^2}{2\Delta}\delta_{kp}\zeta_y -\frac{B_z^2}{2\Delta^2} v k \, \zeta_z \delta_{kp},\qquad
    \hat{I}^{kp} \approx -\frac{e}{L}\frac{v B_z}{\Delta}\zeta_y.
\end{align}
The second term in the effective Hamiltonian describes a shift of the critical (parallel) field from $B_\mathrm{c} = \Delta$ to $B_\mathrm{c} \approx \Delta - B_z^2/2\Delta$. The third term describes the renormalization of the velocity of the Majorana modes due to $B_z \neq 0$. Notice that we retained only the lowest-order term in the gradient expansion of the current operator. This is allowed when $k\xi, p\xi \ll \min (1, m v^2 B_z /\Delta^2 )$. The parameters of the low-energy theory in this case are
\begin{equation}
    v_{R/L} = \Bigl( 1 - \frac{B_z^2}{2\Delta^2} \Bigr) v,\quad g = 0,\quad \alpha \approx \frac{v B_z}{\Delta} . 
\end{equation}
Note that $v_R = v_L$ even though $B_z$ breaks both ${\cal M}_x$ and $\cal{R}$ globally [cf.~Table \ref{tab:class}]. This is a peculiarity of our fine-tuned model with $\mu = 0$. At finite $\mu \ll \Delta$ we find $v_R - v_L \sim v B_z \mu / \Delta^2$, consistently with Table \ref{tab:class}. Finally, we note that in more realistic models of the nanowire,  the orbital effect of the magnetic field might further enhance the mismatch between the velocities of right- and left-moving Majorana modes \cite{nijholt2016}.

\end{document}